\documentclass[twocolumn,secnumarabic,amssymb, nobibnotes,superscriptaddress,showpacs, prb]{revtex4-2}
\usepackage[utf8]{inputenc}
\usepackage{amsmath}
\usepackage{color}
\usepackage{graphicx}
\usepackage{overpic}
\usepackage{caption}
\usepackage{subcaption}
\usepackage{multirow}
\usepackage{enumerate}
\usepackage{bm}
\usepackage{xcolor}
\usepackage{comment} 
\usepackage[normalem]{ulem}

\def\vii{{\bm v}}

\def\BBb{{\bm B}}
\def\ace{\varphi}

\def\cBB{c}

\date{\today}

\begin{document}
\title{Efficient local atomic cluster expansion for BaTiO$_3$ close to equilibrium}
\author{Anna Gr\"unebohm}
\affiliation{Interdisciplinary Centre for Advanced Materials Simulation (ICAMS),  Ruhr-University Bochum, 44780  Bochum, Germany}
\affiliation{Center for Interface-Dominated High Performance Materials (ZGH), Ruhr-University Bochum, 44780  Bochum, Germany}
\author{Matous Mrovec}
\affiliation{Interdisciplinary Centre for Advanced Materials Simulation (ICAMS),  Ruhr-University Bochum, 44780  Bochum, Germany}
\author{Maxim N.\ Popov}
\affiliation{Materials Center Leoben (MCL) Forschung GmbH, A-8700 Leoben, Austria}
\author{Lan-Tien Hsu}
\affiliation{Interdisciplinary Centre for Advanced Materials Simulation (ICAMS),  Ruhr-University Bochum, 44780  Bochum, Germany}
\affiliation{Center for Interface-Dominated High Performance Materials (ZGH), Ruhr-University Bochum, 44780  Bochum, Germany}
\author{Yury Lysogorskiy} 
\affiliation{Interdisciplinary Centre for Advanced Materials Simulation (ICAMS),  Ruhr-University Bochum, 44780  Bochum, Germany}
\author{Anton Bochkarev} 
\affiliation{Interdisciplinary Centre for Advanced Materials Simulation (ICAMS),  Ruhr-University Bochum, 44780  Bochum, Germany}
\author{Ralf Drautz}\affiliation{Interdisciplinary Centre for Advanced Materials Simulation (ICAMS),  Ruhr-University Bochum, 44780  Bochum, Germany}
\affiliation{Center for Interface-Dominated High Performance Materials (ZGH), Ruhr-University Bochum, 44780  Bochum, Germany}

\begin{abstract}
Barium titanate (BTO) is a representative perovskite oxide that undergoes three first-order ferroelectric phase transitions related to exceptional functional properties. In this work, we develop two atomic cluster expansion (ACE) models for BTO to reproduce fundamental properties of bulk as well as defective BTO phases. The two ACE models do not target full transferability but rather aim to examine the influence of implicit and explicit treatment of long-range Coulomb interactions. We demonstrate that both models describe equally well the temperature induced phase transitions as well as polarization switching due to applied electric field. Even though the parametrizations are based on a limited number of configurations that are mostly not far away from the equilibrium, the ACE models are able to capture also properties of important crystal defects, such as oxygen vacancies, stacking faults and domain walls. A systematic comparison shows that the phase transitions as well as the fundamental properties of the investigated defects can be described with similar accuracy with or without explicit treatment of charges and Coulomb interactions allowing for efficient short-range machine learning potentials.
\end{abstract}
\maketitle

\clearpage

\newpage
\section{Introduction}

Ferroelectrics are materials with a switchable spontaneous electric polarization ($\bm{P}$),  which is responsive to 
 electrical fields ($\bm{E}$), strain, thermal energy and light. The resulting functional properties are relevant for applications ranging from memory devices, energy storage and harvesting, condensators and actuators to materials for neuromorphic computing or novel cooling devices \cite{whatmore_100_2021,Smyth2000,Acosta2017,Buscaglia2021,grunebohm_interplay_2021,li_reproducible_2020,Mayer2023}.
Fundamental understanding and optimization of these properties are complex tasks across different scales, since the ferroelectric instability is driven by the electronic structure, while ferroelectric switching and phase transitions are governed by domain wall motion and nucleation at larger scales \cite{grunebohm_interplay_2021}. Furthermore, the latter mechanisms are strongly affected by crystal imperfections, including vacancies and defect dipoles~\cite{ren_large_2004}, dislocations~\cite{hofling_control_2021} and interfaces, as well as local variations in strain and stoichiometry induced by such defects. 

As it is often difficult to  clarify the role of these inhomogeneities and imperfections by experimental means, computer simulations can provide valuable complementary insights \cite{ghosez_modeling_2022,grunebohm_interplay_2021}. On the one hand, 
phenomenological models such as the Ginzburg-Landau-Devonshire \cite{marton_domain_2010} or  Kolmogorov-Avrami-Ishibashi \cite{genenko_multistep_2020} models have been used to analyze the ferroelectric phase stability and switching  on  meso and macro scales. On the other hand, density functional theory (DFT) provides {\emph{ab initio}} access to the atomistic and electronic scales, but its computational expense makes it impractical for sampling temperature variations,  time-dependent material responses or large-scale imperfections.  

To bridge the scales in both time and space, various approximate models have been developed. One widely used approach follows a systematic expansion of the potential energy surface (PES) around a reference configuration either in atomistic \cite{Wojdel2013,escorihuela-sayalero_efficient_2017} or coarse grained degrees of freedom \cite{Zhong1995,Nishimatsu2010,Paul2017}. This approach, either referred to as second principles or effective Hamiltonian  method, has been  very successfully applied to study ferroelectric properties up to the micrometer scale. 
However, the coarse-grained models do not give direct access to atomistic properties or are restricted to specific phonon modes. 

Alternatively, classical interatomic potentials map the chemical interactions on analytic functions. While angular-dependent and many-body terms are commonly neglected for ferroelectric perovskites, although ferroelectricity may compete with rotations of the oxygen octahedra which are sensitive to   angular dependent p-d orbital hybridization \cite{liu_reinterpretation_2013},  the modeling of charges and polarizability is deemed critical. Simple point-charge models can predict structural properties of many oxides surprisingly well and even provide reliable descriptions of extended defects, such as stacking faults and dislocations for some materials \cite{hirel_theoretical_2010}. However, these simple models fail to predict ferroelectric properties. 
The electronic polarizability has been successfully added either by splitting each atom into a core and a shell with 
fractional charges, which can shift relative to each other \cite{cochran_crystal_1960, tinte_ferroelectric_2004} 
or by adding a bond-valence vector that mimics the local symmetry breaking by polar off-centering \cite{liu_reinterpretation_2013}.
These classes of atomistic models have been used successfully to model ferroelectric phase transitions and domain wall (DW) motion \cite{dimou_pinning_2022,vielma_shell_2013,tinte_ferroelectric_2004,liu_reinterpretation_2013,shin_nucleation_2007}.
However, their validity is usually limited and they are rarely transferable to complex defect and structures far from equilibrium \cite{chan_machine_2019}.

In recent years,  machine learning interatomic potentials (MLIPs) have gained great popularity for their accurate description of chemical bonding.  Most MLIPs assume short-sightedness of quantum mechanical interactions and utilize local atomic descriptors with relatively short-range cutoff distances.  This assumption is well founded for metallic and covalent materials, but may be less appropriate for charged systems, where unscreened long-range Coulomb interactions become significant~\cite{behler_four_2021, rinaldi_2025}. Perovskite oxides belong to materials with a mixed type of chemical bonding consisting of covalent and ionic interactions. Recently, several local MLIPs were developed that reproduced specific properties of various ferroelectric perovskites~\cite{jinnouchi_phase_2019,gigli_thermodynamics_2022,gigli_modeling_2024,deguchi_asymmetric_2023,thong_machine_2023,he_structural_2022}. 
Nevertheless, an explicit treatment of long-range interactions, charge transfer or electronic polarizabililty is deemed crucial for a correct description of dipole-dipole interactions, the coupling between polarization and an external electrical field, the depolarization at inhomogeneities and (charged) defects, or the LO-TO splitting of phonon modes~\cite{monacelli_electrostatic_2024}.
Typically, the PES representation and long-range dielectric properties are treated independently, e.g., by additional descriptors for the polarization based on maximally localized Wannier functions \cite{gigli_thermodynamics_2022,zhang_deep_2022,xie_thermal_2024,zhang_deep_2022}
or additional long-range interactions added to the short-range MLIP model \cite{monacelli_electrostatic_2024}. Alternatively, MLIPs were trained on the full electric enthalpy and response~\cite{falletta_unified_2024}.

In this work, we develop two atomic cluster expansion (ACE) \cite{drautz_atomic_2019} models for the prototypical ferroelectric material BaTiO$_3$ (BTO) with the aim to examine the influence of implicit and explicit treatment of long-range Coulomb interactions. Besides its technological importance, BTO undergoes three first-order ferroelectric phase transitions from cubic ($Rm\bar3m$, paraelectric) to tetragonal ($\bm{P}_{\langle100\rangle}$, $P4mm$),  orthorhombic ($\bm{P}_{\langle110\rangle}$, $Amm2$) and rhombohedral  ($\bm{P}_{\langle111\rangle}$, $R3m$) phases with decreasing temperature and present thus a suitable complex system for validation of broad range of properties.
The two models are (i) a solitary ACE parametrization (further referred to as ACE) and (ii) a hybrid model consisting of a local ACE potential and an explicit long-range Coulomb term (further referred to as ACE$^{+}$). 
We show that both potentials are able to describe the complex BTO phase diagram, the character of the phase transitions, field-induced switching as well as several defect structures including oxygen vacancies, stacking faults and domain walls.

\section{Methodology}

\subsection{ACE models}

The ACE methodology combines the advantages of ML methods and physically based models for interatomic interactions. In the following, we summarize the essentials of ACE and refer to original publications for more details \cite{drautz_atomic_2019, drautz_atomic_2020, dusson_atomic_2022, lysogorskiy_performant_2021, bochkarev_efficient_2022, bochkarev_multilayer_2022, bochkarev_graph_2024}. One of the key features of ACE is a complete and hierarchical set of basis functions $\BBb_{i \vii}$ that span the space of local atomic environments. This enables to expand an atomic property $\ace_i^{(p)}$, such as the energy of atom $i$, as
\begin{equation}
\ace_i^{(p)} = \sum_{\vii}^{n_{\vii}} \cBB_{\vii}^{(p)} \BBb_{i \vii} \,, \label{eq:ACEproperty} 
\end{equation}
with expansion coefficients $\cBB_{\vii}^{(p)}$ where ${\vii}$ is composed of several indices.  The basis functions fulfill fundamental translation, rotation, inversion and permutation (TRIP) invariances for the representation of scalar variables, or equivariances for the expansion of vectorial or tensorial quantities. 

Most ACE parametrizations to date use two atomic properties with a Finnis-Sinclair square-root embedding,
\begin{equation}
E_i =  \ace_i^{(1)} + \sqrt{\ace_i^{(2)}}  \,. \label{eq:EFS}
\end{equation}
The square-root non-linearity improves the convergence of the expansion and makes it computationally efficient due to lower number of parameters, compared to a purely linear representation.

In order to test whether long-range Coulomb interactions are only beneficial for the description of ferroelectrics or necessary to model field-induced switching, we compare two different potentials: 
\begin{itemize}
    \item[{ACE~}] The short-range ACE interactions are trained on the full DFT energy, neither long-range interactions of charges nor dipoles are treated explicitly.
    \item[{ACE$^+$}] Analogous to the classical core-shell potentials \cite{tinte_ferroelectric_2004}, long-range interactions between local charges are treated separately from the short-range potential.
    For simplicity, we  use nominal charges (Ba$^{+2}$, Ti$^{+4}$, O$^{-2}$) and subtract the corresponding long-range Coulomb interactions from the DFT-energy before we train the ACE$^+$ model.
\end{itemize}

For both models an additional repulsive term is added for small atomic distances, which have not been considered in the training data, to ensure stability for configurations with a correspondingly large extrapolation grade. As verified for the phase diagram, the additional repulsion does not influence thermodynamic observables, but prevents instabilities, particularly if Ba-Ba and Ti-Ti distances become too short. 

\begin{figure}[ptbhpb]
    \centering
\includegraphics[width=0.5\textwidth]{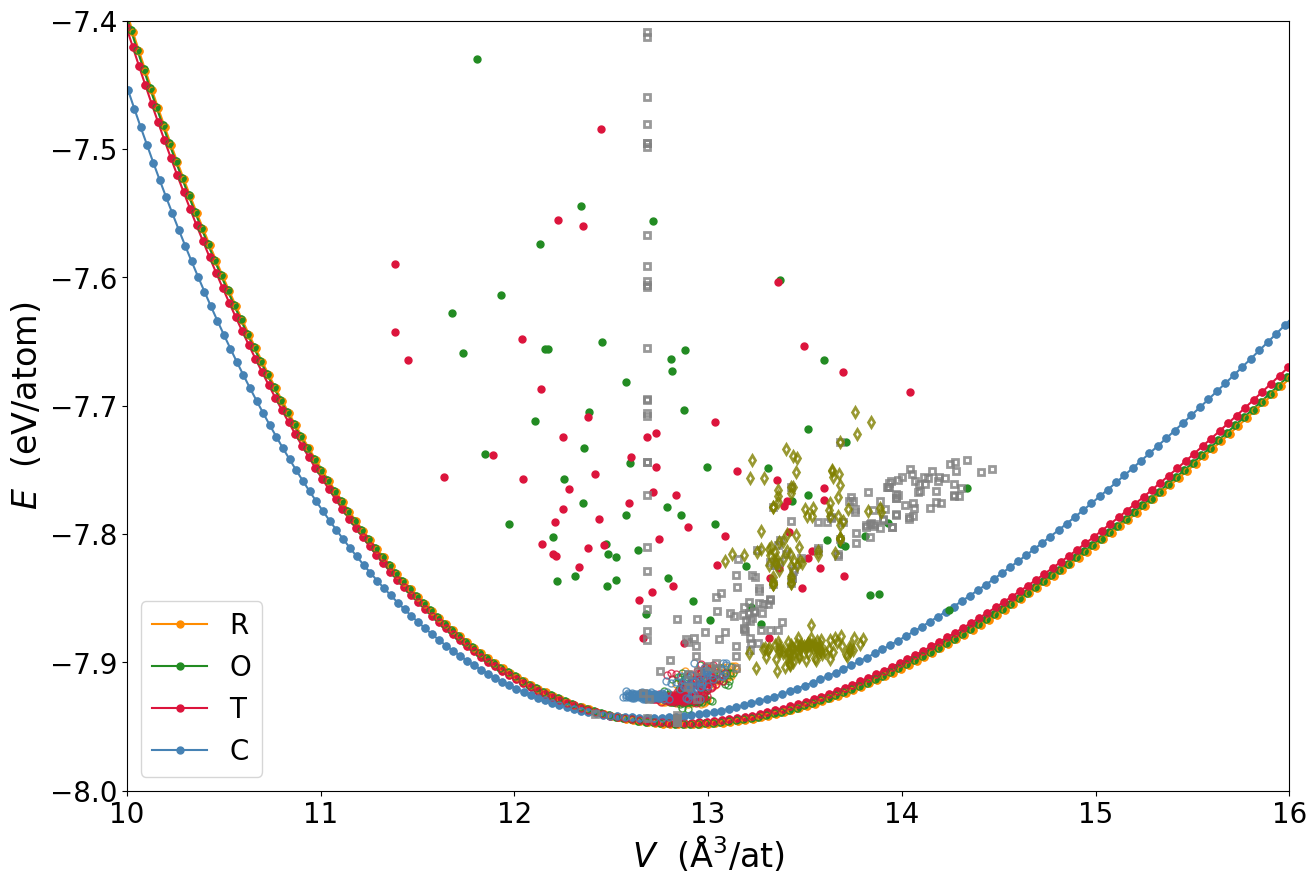}
    \caption{Distribution of DFT training data in volume--energy space. Structures closely related to the four BTO phases cubic (C), tetragonal (T), orthorhombic (O) and rhombohedral (R) are marked by different colors, planar defects and structures from MD simulations are marked as grey squares and olives diamonds, respectively. See main text for details.}
    \label{fig:inputs}
\end{figure}
\subsection{DFT training data}
\begin{table*}[t]
\centering   
\caption{\label{tab:dft} The set of training structures consisted of (i) primitive unit cells of cubic (C), tetragonal (T), orthorhombic (O) and rhomobohedral (R) phases with a broad range of volume and shear deformations, (ii) T180 and T90 (one configuration only) domain wall (DW) configurations, (iii) stacking faults (SF) on the C$\{110\}$ plane, (iv) supercells with randomly displaced atoms, and (v) configurations from molecular dynamics (MD) simulations extracted via active learning.}
\begin{ruledtabular}
\begin{tabular}{llccc}
&{Category} & Description  & No. of structures & No. of atoms in structure \\
\hline 
Unit cells& C, R, O, T  & Volume, and cell shape deformations & 1155 & 5 \\
\multirow{4}{*}{Supercells}& C, R, O, T & Random atomic distortions & 600 & 135 \\
 & T180 and T90 DW & Domain walls & 142 & 20(41), 40(100), 90(1) \\
 & SF  & Stacking faults on the C$\{110\}$ plane  & 198 & 40 \\
 & MD  & Cooling/heating from active learning  & 166 & 40 \\
\end{tabular}
\end{ruledtabular}
\end{table*}

The DFT calculations were performed using the VASP code~\cite{Kresse1993,Kresse1994,Kresse1996,Kresse1996a,Kresse1999} with  exchange-correlation interactions treated at the PBEsol~\cite{Perdew2008} level of theory.
The employed energy cut-off was 520 eV and the Brillouin zone was sampled using uniform grids with $8\times8\times8$ points for the bulk phases and accordingly adjusted grids for larger supercells, in combination with a Gaussian smearing  of 0.01~eV. The electronic self-consistency cycle was run until reaching a tight energy convergence criterion of 10$^{-8}$~eV. Ionic and cell-shape relaxations were conducted so as to reach residual forces not exceeding 10$^{-3}$~eV/\AA.

The total number of training structures was 2261 with the number of atoms in the structures ranging from 5 to 135. The structures included both primitive cell and supercells of the R, O, T and C phases. The primitive cells were subject to a  range of homogeneous volume deformations as well as shear distortions to sample their elastic properties. The supercells served to mimic finite temperature behavior by introducing random displacements of atoms up to 0.2~\AA. A limited number of bulk defects were included as well, namely, 180$^{\circ}$ and 90$^{\circ}$ (one configuration only) domain walls 
in the tetragonal phase and generalized stacking faults (SF) on the $\{110\}$ plane in the cubic phase. Finally, a small fraction of the training configurations was obtained using active learning. For this purpose, an initial ACE parametrization, obtained using the configurations described above, was used in MD simulations, where the cubic phase was first gradually cooled down from 375 to 125~K and then heated up again. Additionally, for the tetragonal and cubic phases, we conducted $NPT$ simulations at $P=-2$~GPa over a temperature range of 300?400~K. In all these MD runs, we monitored the extrapolation grade $\gamma$ based on the D-optimality criterion~\cite{lysogorskiy2023active} and added configurations with large values of $\gamma$ to the training data. 

A summary of training structures and the number of structures in each category are provided in Table~\ref{tab:dft}. As shown in Fig.~\ref{fig:inputs},  most training structures are confined in a narrow volume region around the equilibrium volume to capture accurately small energy differences between different ferroelectric phases.  We stress that the chosen moderate number of configurations does not guarantee transferability of the present ACE parametrizations to configurations with large local atomic distortions, configurations with large inhomogeneous deformations or density changes, and defective structures such as surfaces or interfaces with altered stoichiometry. In simulations under such conditions, the extrapolation grade needs to be closely monitored and upfitting with additional DFT simulations may be necessary.

\subsection{Technical details}
The simulations were conducted using the Performant Implementation of the Atomic Cluster Expansion (PACE) \cite{lysogorskiy_performant_2021} as implemented in the Atomic Simulation Environment (ASE)~\cite{ase-paper} and the LAMMPS package \cite{plimpton_fast_1995,lammps}. 
The cutoff of both ACE potentials was set to 6~\AA.  In case of ACE$^{+}$, the long-range Coulomb interactions were treated with the Ewald method and  we used  the cutoff of  7~\AA\, for the real part of the Ewald summation, beyond which the interactions are treated in reciprocal space. This choice leads to convergence of the Coulomb energy within $10^{-5}$ eV/atom.

An initial validation of  fundamental properties (geometry, elastic moduli, phonons) of the perfect bulk phases was performed using ASE libraries~\cite{ase-paper}.  Thresholds for maximum residual forces and stresses for relaxations were set to 0.01 eV/\AA~and 50 MPa.
Larger systems were evaluated using LAMMPS. In static simulations, cell shape, volume and atomic positions were optimized without any imposed symmetry relation  until the energy change between successive iterations divided by the energy magnitude was less than or equal to 10$^{-5}$~eV. 

In molecular dynamics (MD) simulations, the Nos\'e-Hoover thermostat (barostat, 1~bar) was used with a damping parameter of 50~fs  (250~fs), together with a  timestep of 0.5~fs. If not stated otherwise $10\times 10 \times 10$ unit cells were combined with periodic boundary conditions. 
Phase diagrams were evaluated based on heating and successive cooling simulations with a temperature change rate of 1~K/ps. Moving averages  over 5~ps  of lattice parameters and polarization are based on 
output written  every 0.05~ps and 0.25~ps, respectively. The given transition temperatures correspond to the starting temperatures of the transition determined by the change of the lattice parameters.


The local polarization of each Ti centered unit cell at $\bm{r}_{\kappa}^\text{Ti}$ was approximated based on the positions of its six and 8 nearest  O and Ba   neighbours ($\bm{r}_j$) as \cite{sepliarsky_first-principles_2011,he_structural_2022} 
\begin{equation}
{\bm{P}}_{\kappa}=\frac{1}{\Omega_{\kappa}} \sum_{j} w_j \rho_{j} ( \bm{r}_{j} - \bm{r}_{\kappa}^\text{Ti} )\, ,
\label{eq:P}   
\end{equation}
where $\Omega_{\kappa}$ is the unit cell volume determined from the mean Ba--Ba distances, $\rho_j$ is the nominal charge of atom $j$, and $w_j$ is its weight factor, i.e. the reciprocal of the number of cells sharing atom $j$.  
This approach correctly predicts the absence of polarization in centro-symmetric cells and the qualitative changes of polarization with ionic off-centering, but it  underestimates the magnitude of polarization as the electronic polarizability is not included.  
We used Ovito~\cite{ovito} to visualize the local polarization of each unit cell.

The field hysteresis was simulated within an electric field range of $\pm$250~kV/cm, changing the field in 10~kV/cm steps for $\pm$200~kV/cm. At each field strength 20~ps and 5~ps were used for thermalization and recording of thermal averages, respectively.
 While the applied field can directly couple to the atomic charges in case of the ACE$^+$ model, for the ACE model the external electric field (${\text E}_{x}$) was approximated by fictive forces on atoms according to Ref.~\onlinecite{Fu_Bellaiche_2002}:
\begin{equation}
    \textbf{F}_{i}= \rho_{i} \cdot {\bm{E}_{\text{ext}}}.
    \label{eq:F}   
\end{equation}

\begin{figure}
\centering
    \includegraphics[width=0.4\textwidth]{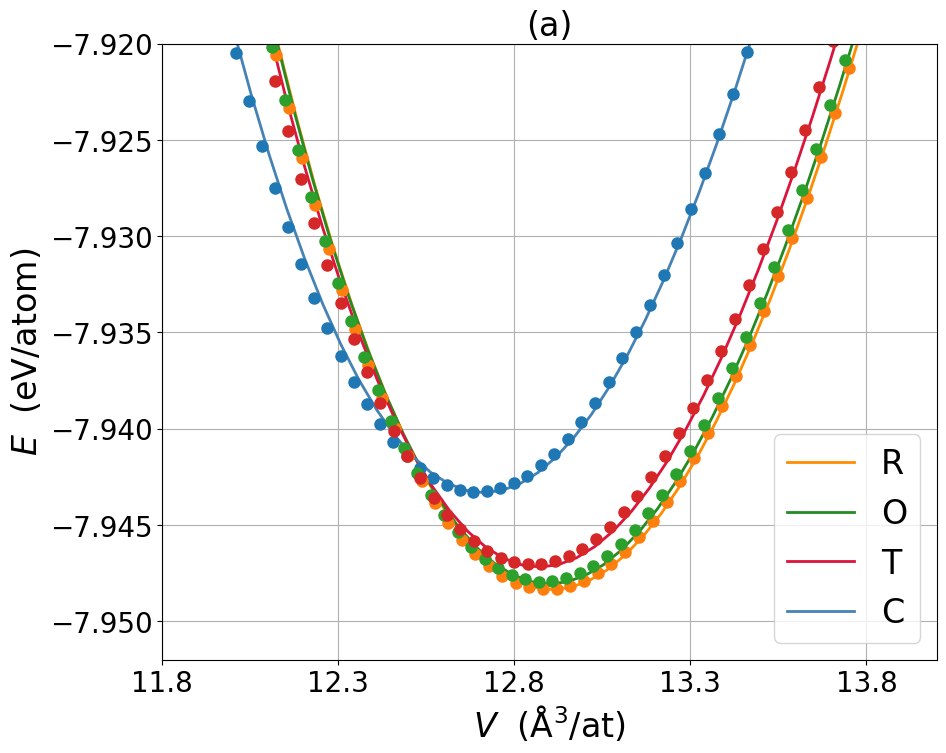}
    \\
    \vspace{0.5cm}
    \includegraphics[width=0.4\textwidth]{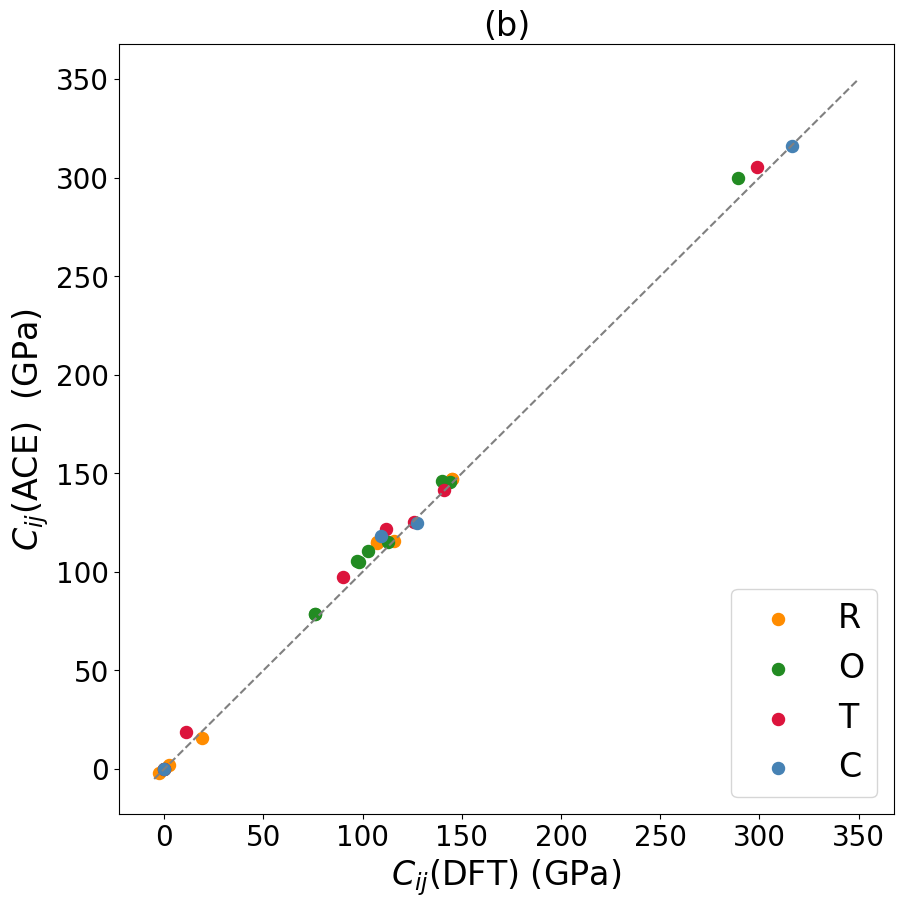}
    \caption{ACE predictions of properties of bulk BTO phases: (a) Energy-volume curves (lines -- ACE, symbols -- reference DFT values); (b) Components of the elastic tensor $C_{ij}$.
    \label{fig:EV-elastic}}
\end{figure}

\section{Ideal bulk phases}
\subsection{Phase stability at 0 K}

The key prerequisite for a reliable description of ferroelectric behavior of BTO is the ability to capture accurately the subtle energy differences between the ground-state configurations of the four equilibrium bulk phases as well as their elastic and vibrational properties.  It should be emphasized that the ground states of the three lowest BTO phases (R, O and T) are only 1~meV/atom apart from each other. For brevity, we present here results of the ACE parametrization only. Predictions of the ACE$^+$ model are equivalent unless noted otherwise and are presented in the Supplementary Material~\cite{supp}.
Figure~\ref{fig:EV-elastic}(a) shows that ACE reproduces the $E(V)$ curves for all four BTO phases around equilibrium in excellent agreement with the reference DFT calculations. Furthermore, as shown in Fig.~\ref{fig:EV-elastic}(b) also the predicted elastic moduli of all phases are in close agreement with the DFT reference data (exact numerical values are provided in the Supplementary Material~\cite{supp}).

\begin{figure}
    \centering
    \centerline{
    \subcaptionbox{\label{f1} $2 \times 2 \times 2$ cubic supercell}[0.25\textwidth]
    {\includegraphics[width=0.25\textwidth,clip,trim=0cm 0cm 0cm .7cm]{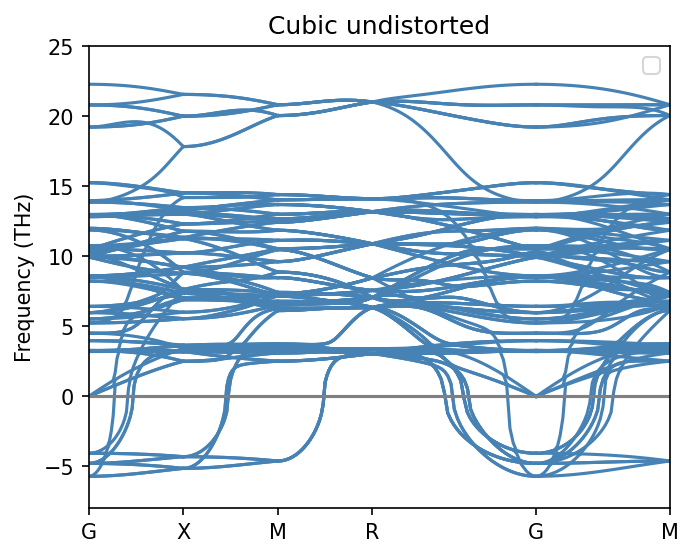}}
        \vspace{0.5cm}
    \subcaptionbox{\label{f2} 2+6 structure}[0.25\textwidth]
    {\includegraphics[width=0.25\textwidth,clip,trim=0cm 0cm 0cm .7cm]{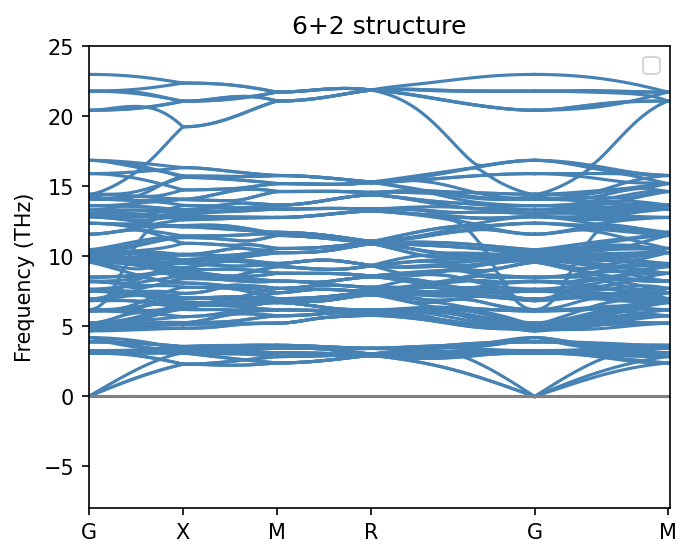}}}
    \subcaptionbox{\label{phon}}[0.5\textwidth]
    {\includegraphics[width=0.45\textwidth]{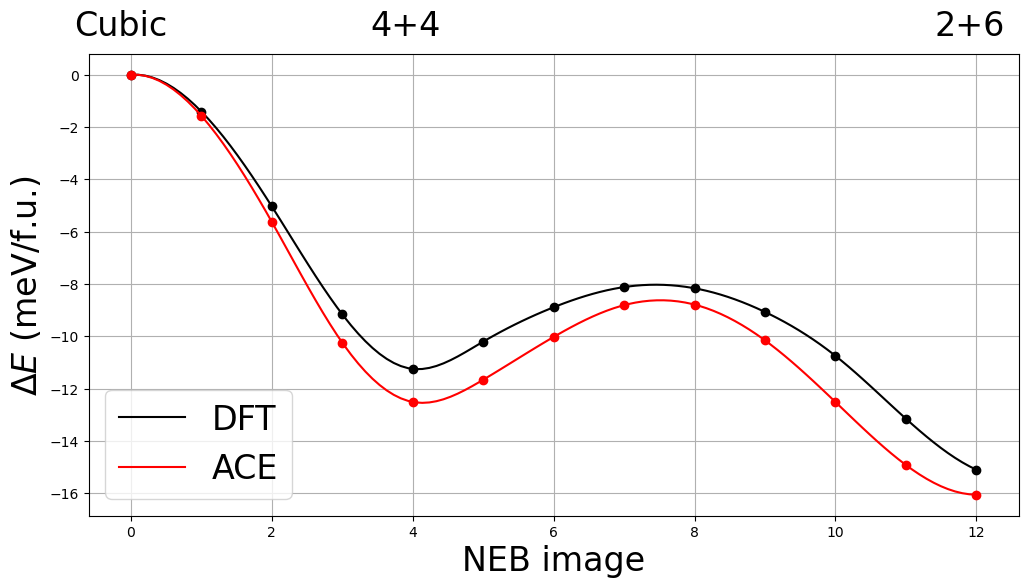}}
        \vspace{-0.5cm}
    \caption{Prediction of phonon properties by ACE: (a) Phonon spectra of the cubic phase showing the expected  unstable phonon modes in large parts of the Brillouin zone. (b) Dynamical stable spectrum of the cubic $2+6$ superstructure predicted in Ref.~\onlinecite{kotiuga_microscopic_2022}.  (a)--(b) Below 15~THz both spectra are in excellent agreement to the previous DFT prediction from \cite{kotiuga_microscopic_2022}. Note that results for $2\times 2 \times 2$ supercells without backfolding and without LO-TO splitting are presented. (c) 
    Potential energy landscape between the perfect cubic phase and the $4+4$ and $2+6$ superstructures predicted by DFT in Ref.~\onlinecite{kotiuga_microscopic_2022} (black) compared to ACE (red).
      \label{fig:kotiuga}}
\end{figure}

Since the phase stability and phase transitions in perovskite oxides depend on the softening of particular phonon modes, a correct description of the low-energy part of the phonon spectrum is essential for any atomistic potential. 
As shown in Fig.~\ref{fig:kotiuga}(a), the perfect undistorted cubic phase exhibits three unstable phonon modes at the $\Gamma$, $M$ and $X$ points of the Brillouin zone.
All three instabilities are correctly captured by ACE, even though they were not explicitly included in the training data. Particularly, the $\Gamma$-centered mode related to the ferroelectric phase transition is lowest in energy as expected. 
The frequencies of the soft modes are in better quantitative agreement with DFT than for other ML models~\cite{gigli_thermodynamics_2022}.
Furthermore, all phonon branches below 15~THz are reproduced in close agreement with DFT~\cite{kotiuga_microscopic_2022}.

Since the LO-TO splitting at $\Gamma$ is not included in the DFT training data, the high-energy modes around $\Gamma$ are not predicted correctly by ACE, which limits the accuracy of the potential at elevated temperatures and large excitation. It has been shown recently \cite{monacelli_electrostatic_2024} that an 
approach similar to the Wolf-summation technique \cite{wolf_exact_1999} can be employed to correct the LO-TO splitting, which can be added to the present ACE implementation in the future.

To validate the ACE predictions of phonon instabilities in more detail, we followed a recent work by Kotiuga et al.~\cite{kotiuga_microscopic_2022} who  identified two structural prototypes with $I\bar{4}3m$ and
$Pa\bar{3}$ symmetries, further referred to as $4+4$ and $2+6$ structures, that remove the dynamic instabilities in the cubic phase. 
Both structures can be constructed from $2 \times 2 \times 2 $ supercells of the five-atom primitive cubic cell by off-centering the Ti positions. Since there exist unique transformations between the cubic and $4+4$ structures as well as the $4+4$ and $2+6$ structures, it is possible to obtain the minimum energy pathway characterizing the transformation using the nudged elastic band method \cite{henkelman_climbing_2000}. The transformations computed by ACE and DFT~\cite{kotiuga_microscopic_2022} are presented in Fig.~\ref{fig:kotiuga}.  It is obvious that ACE is not only able to reproduce the energy differences between the structures but also the energy barriers. Furthermore, the unstable phonon modes in the cubic supercell indeed disappear in the $4+4$ (not shown) and $2+6$ structures (Fig.~\ref{f2}) in DFT and ACE. 

\begin{figure}
    \centering
    \includegraphics[width=0.5\textwidth]{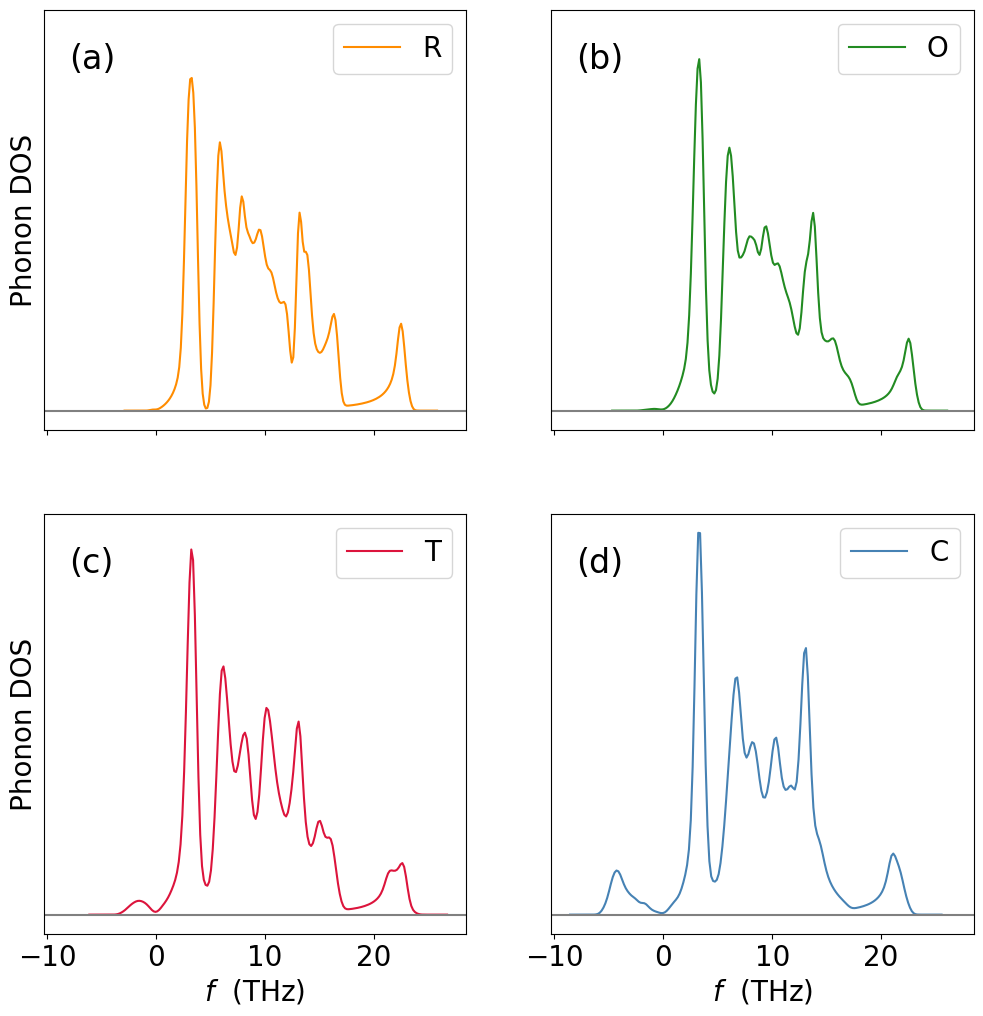}
    \caption{
    Phonon densities of states at 0~K (primitive cells, no LO-TO splitting) as predicted by the ACE potential 
    for (a) rhombohedral, (b) orthorhombic, (c) tetragonal, and (d) cubic phases of BTO.}
    \label{fig:phonons}
\end{figure}

Without large scale displacement patterns like $4+4$ and $2+6$, C, T, and O phases of BTO are dynamically unstable and going from C to O phases, the weight of these modes in the phonon DOS decreases systematically. As shown in Figs.~\ref{fig:phonons}(a)-(d), ACE predicts all these trends correctly.

\subsection{Finite temperatures and phase transitions}
\begin{figure}[tpbh]
    \centering
    \begin{overpic}[width=0.45\textwidth]{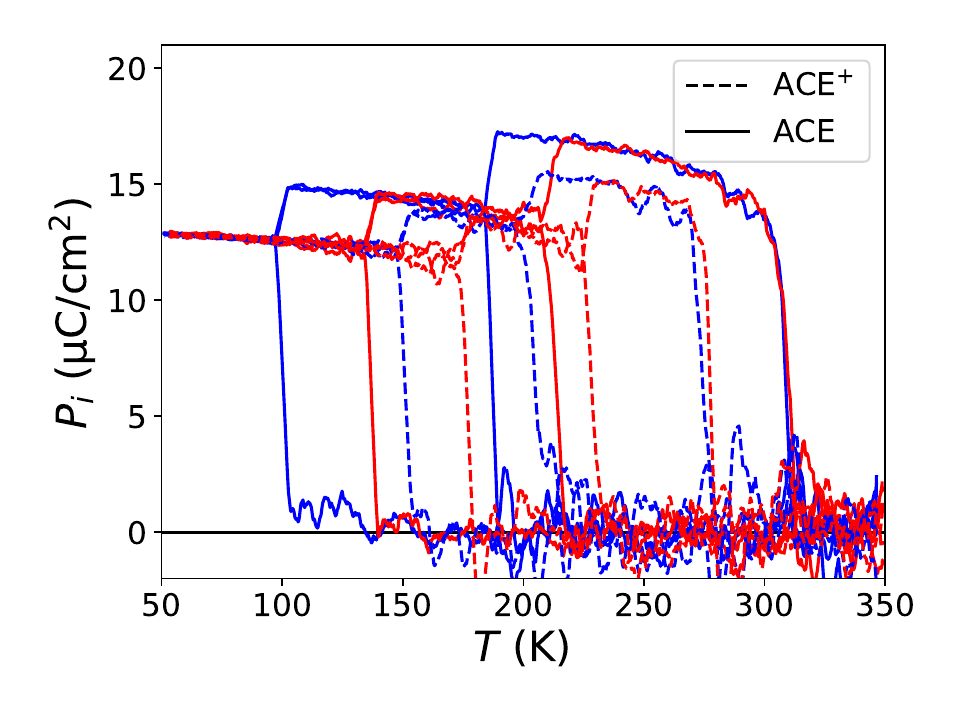}    
\put(0,65){\text{\Large(a)}}
 \end{overpic}
    \begin{overpic}[width=0.45\textwidth]{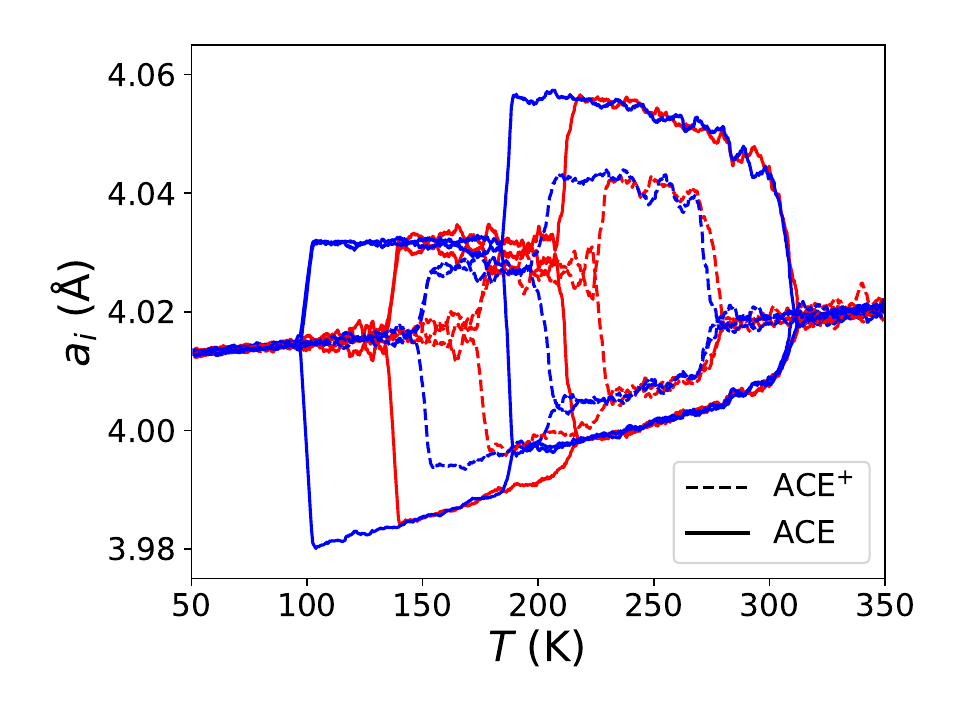}  
\put(0,65){\text{\Large(b)}}
 \end{overpic}
    \caption{Changes of the three (a) polarization components $P_i$ and (b) lattice parameters $a_i$  observed during cooling (blue) and heating (red) simulations using ACE (solid lines) and ACE$^+$ (dashed lines) potentials.}
    \label{fig:phase_diagram}
\end{figure}

\begin{table}[thpb]
\centering
\caption{\label{tab:Tc}Comparison of phase transition temperatures ($T_C$) obtained by simulations and experiment. If $T_C$ has been reported for cooling and heating the mean value is given.  The three columns on methods give the type of the potential, the exchange-correlation functional of the DFT calculations, and corrections and modifications. 
}
\begin{tabular}{|c|c|c|ccc|}
\hline
\multirow{2}{*}{Method} &\multirow{2}{*}{Data set} &\multirow{2}{*}{Correction} &\multicolumn{3}{c|}{$T_C$ (K)}                           \\ 
     &                  &   & \multicolumn{1}{c|}{R--O} & \multicolumn{1}{c|}{O--T} & T--C \\ \hline
ACE~~     & PBEsol  & -&\multicolumn{1}{c|}{119}    & \multicolumn{1}{c|}{199}& \multicolumn{1}{c|}{308}   \\ \hline
ACE$^+$ & PBEsol  & - &\multicolumn{1}{c|}{163}    &  \multicolumn{1}{c|}{217}  & \multicolumn{1}{c|}{274}    \\ 
\hline\hline
\multirow{6}{*}{${\cal{H}}_{\text{eff}}$}&  LDA\footnotemark[1]      &-& \multicolumn{1}{c|}{95}  & \multicolumn{1}{c|}{110} & 137 \\ 
&  WC GGA\footnotemark[1]   &- &\multicolumn{1}{c|}{102} & \multicolumn{1}{c|}{160} & 288 \\ 
&  SCAN\footnotemark[2]     & - & \multicolumn{1}{c|}{111} & \multicolumn{1}{c|}{141} & 213 \\ 
 &  SCAN\footnotemark[2] &Anharm. coupl.           & \multicolumn{1}{c|}{230} & \multicolumn{1}{c|}{278} & 375 \\ 
&  PBEsol\footnotemark[3] &  Anharm. coupl.        & \multicolumn{1}{c|}{186} & \multicolumn{1}{c|}{255} & 395 \\ 
& PBEsol\footnotemark[4]& Anharm. coupl.& \multicolumn{1}{c|}{220} & \multicolumn{1}{c|}{270} & 380\\
\hline
\multirow{3}{*}{Core-Shell} & LDA\footnotemark[5]&PES rescaled       & \multicolumn{1}{c|}{140} & \multicolumn{1}{c|}{190} & 360 \\ 
   & PBEsol\footnotemark[6]&PES rescaled    & \multicolumn{1}{c|}{180} & \multicolumn{1}{c|}{250} & 340 \\
  & PBEsol\footnotemark[6]          & -&\multicolumn{1}{c|}{150} & \multicolumn{1}{c|}{210} & 260 \\ 
 \hline
Allegro & PBEsol\footnotemark[7]&- & \multicolumn{1}{c|}{$\approx$60} &\multicolumn{1}{c|}{$\approx$200}& \multicolumn{1}{c|}{$\approx$260} \\ 
  SOAP-GAP& PBEsol\footnotemark[8] &-& \multicolumn{1}{c|}{19}&\multicolumn{1}{c|}{91}&\multicolumn{1}{c|}{182}\\ 
 SOAP-GAP&r$^2$SCAN\footnotemark[9]& -&\multicolumn{1}{|c|}{-}&-&\multicolumn{1}{|c|}{254}\\
\hline
\multicolumn{3}{|c|}{Exp.\footnotemark[10]}                                           & \multicolumn{1}{c|}{183} &\multicolumn{1}{c|}{278} & 403 \\ \hline
\end{tabular}\\
\footnotemark[1]Ref.~\onlinecite{Nishimatsu2010}, \footnotemark[2]Ref.~\onlinecite{Paul2017}, \footnotemark[3]Ref.~\onlinecite{Mayer2022}, \footnotemark[4]Ref.~\onlinecite{ma_active_2025},  
\footnotemark[5]Ref.~\onlinecite{tinte_ferroelectric_2004}, \footnotemark[6]Ref.~\onlinecite{vielma_shell_2013}, \footnotemark[7]Ref.~\onlinecite{deguchi_asymmetric_2023}, \footnotemark[8]Ref.~\onlinecite{gigli_thermodynamics_2022},
\footnotemark[9]Ref.~\onlinecite{gigli_modeling_2024},
\footnotemark[10]Ref.~\onlinecite{kay_smmetry_1949}.
\footnotemark[8]\footnotemark[9]Transition temperatures estimated from free energies.
\end{table}
The correct description of energetics at zero temperature does not guarantee correct phase transformations at finite temperatures. A series of MD simulations using both ACE and ACE$^+$ models were conducted to demonstrate correct finite temperature behavior. Figure~\ref{fig:phase_diagram} shows the observed changes of polarization and lattice parameters with temperature from heating and cooling MD runs.  Both ACE models give the expected sequence of phase transformations in BTO, namely, C $\rightarrow$ T $\rightarrow$ O$\rightarrow$ R with decreasing temperature.
The total polarization, computed using Eq.~\eqref{eq:P}, amounts to about 17, 21 and 23~$\mu$C/cm$^2$ for ACE and 15, 20 and 23~$\mu$C/cm$^2$ for ACE$^+$, respectively. These values are somewhat lower than experimental values of 18, 27 and 33~$\mu$C/cm$^2$~\cite{menoret_structural_2002}, but qualitative trends are well predicted. Particularly, the polarization directions ($\langle$001$\rangle$,$\langle$011$\rangle$ and $\langle$111$\rangle$) are correctly given. Quantitative deviation is  expected as electronic polarizability is not accounted for in the MLIPs. 
Furthermore, both potentials predict identical lattice parameters for the R and C phases. A slight systematic overestimation of the lattice parameters can be related to the exchange-correlation functional employed in our DFT calculations.
In the tetragonal and orthorhombic phases, the predictions of lattice parameters by ACE are 
in better agreement with experiments than those by ACE$^+$ (the $c/a$ ratio is 1.014 for ACE, 1.007 for ACE$^+$ and 1.02 from experiment~\cite{menoret_structural_2002}).

The phase transition temperatures ($T_C$) obtained from our simulations are compared to other theoretical predictions and experimental observations in Tab.~\ref{tab:Tc}. The influence of finite size effects on $T_C$ is shown in the Supplementary Material~\cite{supp}. Note that a $10\times 10 \times 10$ system size was sufficient to model all phase transitions. 
Both our PBEsol-based models underestimate the absolute transition temperatures, but give consistently better results than other existing DFT based models (with a range of 137--288~K for the C--T transition). For example, a recent SOAP-GAP potential~\cite{gigli_thermodynamics_2022} predicts transition temperatures of 19, 91 and 182~K, which may be related to a poor description of soft phonon modes. Note that better agreement with experimental transition temperatures for some core-shell potentials was achieved only by ad-hoc rescaling of the PES. The transition temperatures predicted by the ${\cal H}_{\text{eff}}$ approaches lie also below 300~K in its original formulation. However, it has been reported recently that anharmonic coupling terms in the energy expansion lead to improved predictions \cite{Mayer2022,ma_active_2025}. 

\begin{figure}[thbp]
    \centering
    \includegraphics[trim=4cm 1cm 6cm 1.5cm,clip,width=0.5\textwidth]{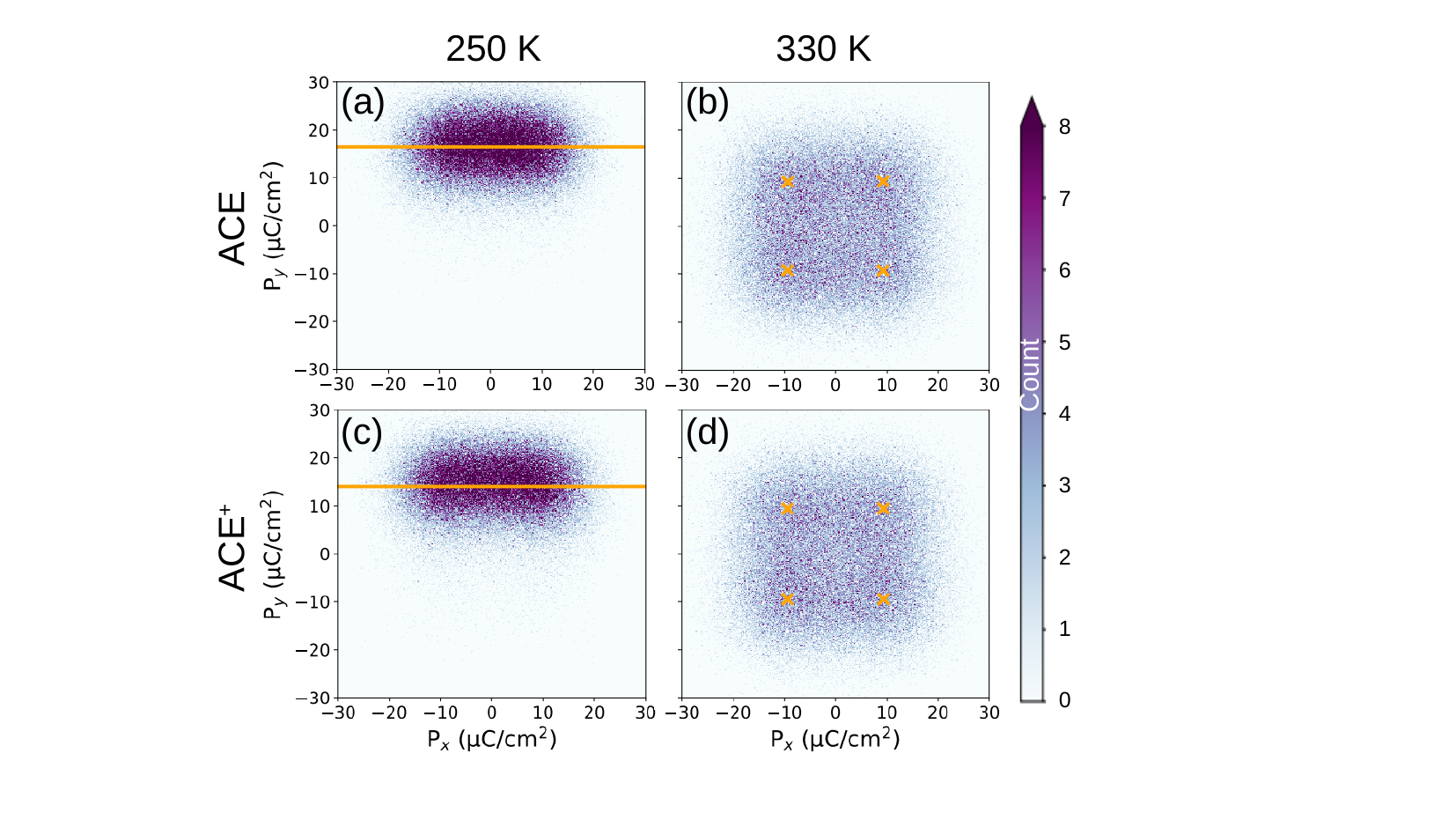}
    \caption{Cumulative polarization distribution based on 100 snapshots (25~ps during cooling simulation) in $P_x-P_y$ space (a, c) in the tetragonal phase at 250~K  and (b, d) in the cubic phase at 330~K obtained by (a)--(b) ACE and  (c)--(d) ACE$^+$. Orange lines and crosses mark the mean values and the mean values in each quadrant, respectively. }
    \label{fig:pdist_xy}
\end{figure}
\begin{figure}[tbh]
    \centering
    \begin{overpic}[trim= 2cm 0.5cm 6cm 0.5cm, clip, width=0.5\textwidth]{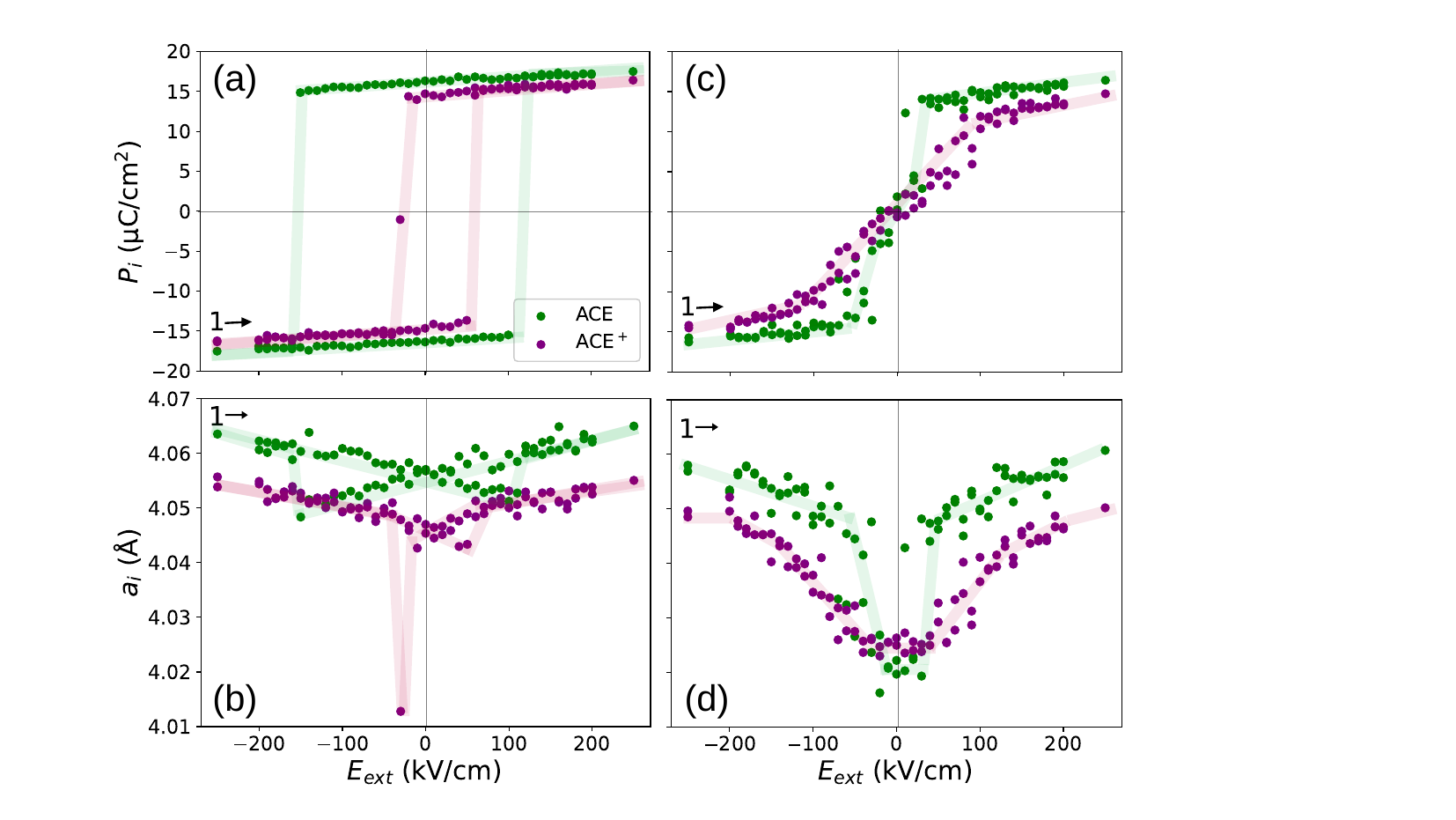}
 \end{overpic}
    \caption{Field hysteresis  (a, b) in the T phase at 250~K  and (c, d) in the C phase at 310~K. Panels (a, c) and (b, d) show the field-induced changes of polarization $P_i$ and lattice parameter $a_i$ parallel to the direction of the applied field, respectively.
    Green curves are for ACE and purple curves are for ACE$^+$. The initial condition of the simulations is marked by "1". Lines are added to guide the eye. Note that the coercive field is met exactly by the chosen sampling only for ACE$^+$ and negative field direction. 
 }
    \label{fig:P_E}
\end{figure}

Not only the sequence of structural--ferroic phase transitions, but also their character is correctly described by both ACEs. Experimentally,  weak first-order transitions with thermal hysteresis of about 2~K~\cite{moya_giant_2013},  10~K~\cite{limboeck_evolution_2014} and 20~K~\cite{von_hippel_ferroelectricity_1950} have been reported for the C $\leftrightarrow$ T $\leftrightarrow$ O $\leftrightarrow$ R transitions. In our simulations, we find abrupt transitions with  apparent thermal hysteresis due to overheating/undercooling of 18~K and 31~K for ACE and 21~K and 24~K for ACE$^+$ for O $\leftrightarrow$ T and R $\leftrightarrow$ O, respectively. The overestimation of the thermal hysteresis of these lower transitions is related to the so-called Landauer paradox \cite{landauer_electrostatic_1957}, i.e.,  missing nucleation centers in ideal systems without any defects. While we also find an abrupt T$\leftrightarrow$C transition, its small thermal hysteresis is subject to finite size effects, cf. Supplementary Material~\cite{supp}.

Furthermore, both ACE models correctly predict the mixed displacive and order-disorder character of the transitions \cite{kotiuga_microscopic_2022,Comes_1968,cochran_crystal_1960,gigli_thermodynamics_2022} as can be seen by the distributions of local dipoles in $P_x$--$P_y$ space in T and C phases in Fig.~\ref{fig:pdist_xy}. Similar outcomes emerge for distributions in the $z$-direction and for the other transitions.  Already at 330~K, i.e., well in the C phase, the local dipoles are not zero, but show a broad distribution around the eight $\langle 111 \rangle$ directions. At the transition, the dipoles order readily along the tetragonal direction (polarization along $+y$ in Fig.~\ref{fig:pdist_xy}) and at the same time the mean value of $p_y$  increases. 
Since there are no significant difference between ACE and ACE$^+$, the influence of explicit long-range Coulomb interactions seems to be irrelevant for the appearance and character of the phase transitions.

\subsection{Polarization switching}

Polarization switching and coupling between polarization and applied field is most relevant for practical applications.  As described above, electric fields can be applied directly only for ACE$^+$, where atoms possess explicit charges. In the case of ACE, the electric field is applied indirectly by exerting forces on atoms according to Eq.~\eqref{eq:F}.

\begin{figure}[thb]
\centering
\begin{overpic}[trim= 5.8cm 9cm 13.5cm 0.4cm, clip, angle=180, width=0.3\textwidth]{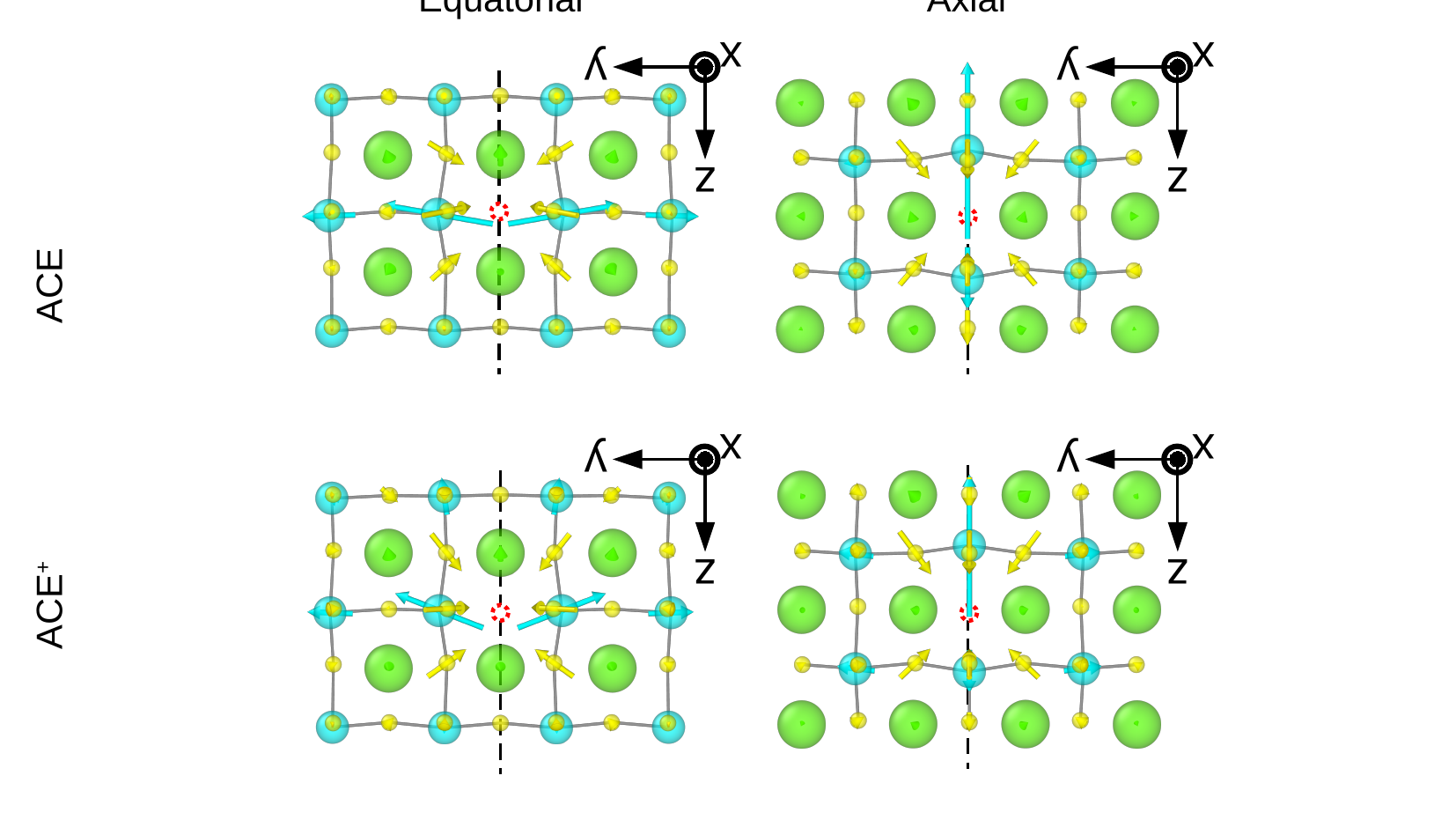}
\put(-9,55){\text{\large(a)}}
 \end{overpic}
\begin{overpic}[trim= 14.8cm 8cm 4.4cm 0.4cm, clip,angle=180, width=0.3\textwidth]{oxygenvacancy.pdf}
\put(-9,55){\text{\large(b)}}
 \end{overpic}
    \caption{Distortions of the atomic structure induced by oxygen vacancies in the tetragonal phase (the tetragonal axis lies along $z$) of BTO as predicted by ACE at 0~K; (a) the equatorial vacancy connecting Ti atoms perpendicular to the polarization direction and (b) the apical (or axial) vacancy connecting Ti atoms along the polarization direction.
Blue, green and yellow spheres represent Ti, Ba and O atoms, respectively. Arrows mark the direction and magnitude of atomic relaxation compared to the pristine material.} 
    \label{fig:point}
\end{figure}

Both models correctly reproduce squared-shaped $P(E)$ hysteresis of an ideal defect-free tetragonal phase, see Fig.~\ref{fig:P_E}(a). At 250~K, the coercive field of ACE is 130~kV/cm, and thus of the same order of magnitude as 260~kV/cm found by the effective Hamiltonian approach (using feram code \cite{Nishimatsu2010} together with the effective Hamiltonian by Zhong et al.\ \cite{Zhong1995} parametrized using DFT calculations for BTO~\cite{Nishimatsu2010}). Furthermore, both ACE and ACE$^+$ qualitatively predict the expected piezoelectric response, namely, a butterfly shape of the $a(E)$ curves in Fig.~\ref{fig:P_E}(b) with a sharp jump of the strain at the coercive field.  However, due to the stochastic nature of the transition and thermal fluctuations, the exact value of the coercive field with $P=0$ is difficult to pinpoint in the simulations. The coercive field was accidentally met only in one case for the negative field applied in ACE$^+$ simulations (the sole point at about $-40$~kV/cm in Fig.~\ref{fig:P_E}(b)). In all other cases the coercive field was not met and hence the minimal lattice parameter at the coercive field is not visible. As shown in Figs.~\ref{fig:P_E}~(c, d), also the field-induced paraelectric-to-ferroelectric transition
\cite{Merz1949} above $T_C$ is well reproduced by both potentials.

These tests show that there is no clear benefit of the explicit treatment of charges and long-range interactions for the prediction of field-induced homogeneous switching and the field induced polarization changes can be addressed comparably well by the ACE model, albeit indirectly.
The main effect of adding explicit charges in ACE$^+$ is a reduction of the coercive field at a given temperature (to about 40~kV/cm at 250~K).  This can be related to a smaller difference in free energies between the tetragonal and cubic transient states at the coercive field in case of ACE$^+$.  ACE$^+$ also predicts smaller $T_C(T-C)$ of 274~K (temperature with equal free energies), which is closer to the sampled temperature of 250~K.  


\section{Defects}
\begin{figure}[tpbh]
    \centering
\includegraphics[width=0.5\textwidth]{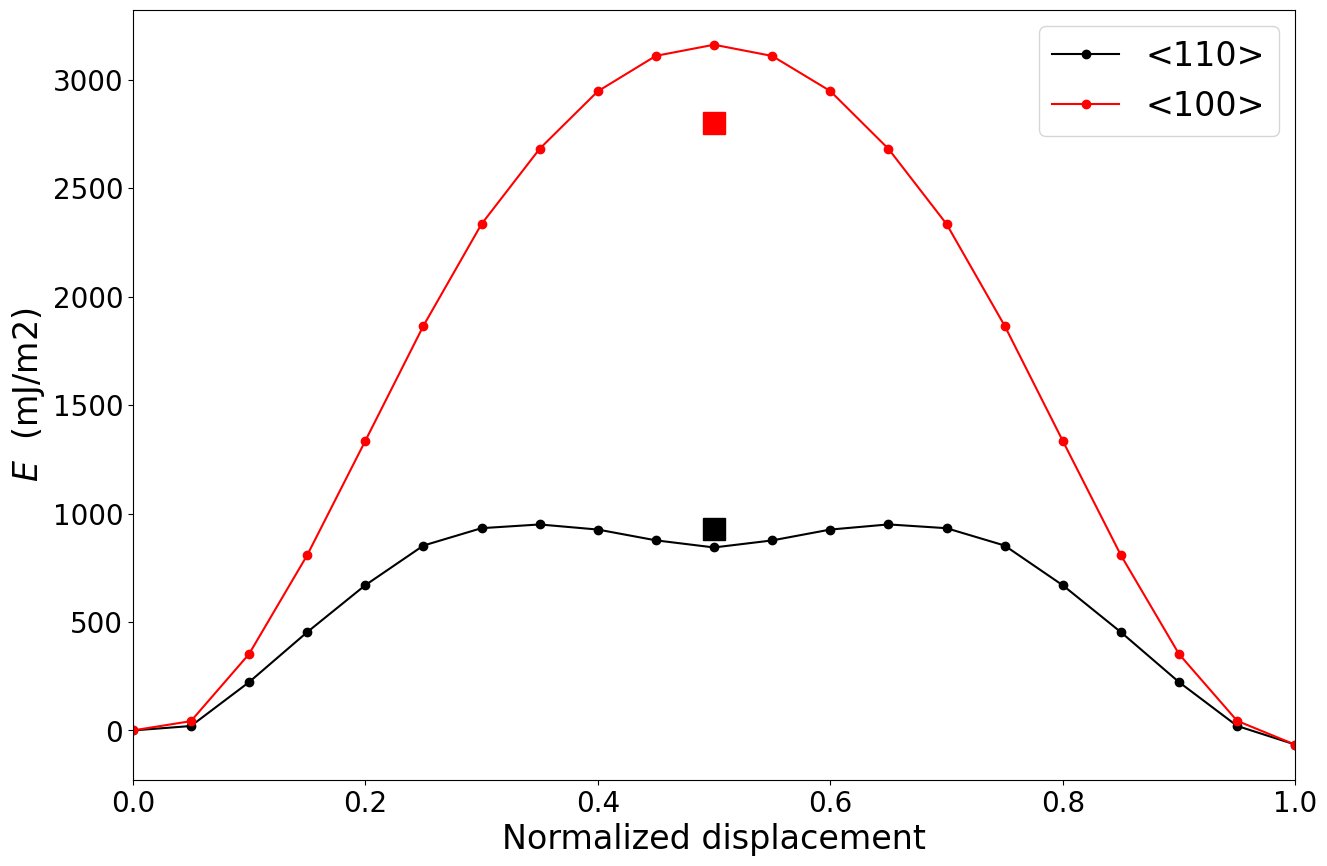}
    \caption{Energy profiles along the $\langle 110 \rangle$ and $\langle 100 \rangle$ directions of the $\{110\}$ $\gamma$-surface computed by ACE (cubic phase). The square symbols are APB energies obtained by  DFT (PBE potential)~\cite{hirel_theoretical_2010}.}
    \label{fig:gs}
\end{figure}

An important advantage of the interatomic potentials over the coarse-grained effective Hamiltonian models is their ability to predict the properties of lattice defects with atomistic resolution. On the one hand, the correct description of defects is challenging for simple  empirical potentials. Here MLIPs can fill important blind spots in the fundamental understanding of defects and their interactions. On the other hand, transferability needs to be verified for all potentials.
In this work, we  examine three prototypical classes of defects: (i) oxygen vacancies, as example for simple point defects, (ii) anti-phase boundaries, as precursor for studies of dislocations, and (iii) domain walls, as the most common 2D defects in ferroelectrics that govern ferroelectric switching and functional properties \cite{grunebohm_interplay_2021}.

Results for oxygen vacancies in the tetragonal phase are illustrated in Fig.~\ref{fig:point}. The corresponding results for ACE$^+$ are shown in the Supplementary Material~\cite{supp}. Due to symmetry of the T phase, there exist two non-equivalent oxygen vacancy sites. The first type is an equatorial oxygen vacancy between Ti atoms perpendicular to the  polarization direction. The second type is an axial vacancy located between Ti atoms along the polarization direction. 
The equatorial vacancy has a higher energy than the apical one both for ACE (61~meV) and ACE$^+$ (108~meV) potentials.  This is in qualitative agreement with the DFT predictions~\cite{rusevich_ab_2020} and can be related to the reduced distance between Ti and the apical O in the ferroelectric phase which induces short-range repulsion. The atomic relaxations around the vacancies are mostly confined to the first neighbor shells of atoms.  Both nearest and second nearest O neighbours shift towards the vacancy while the Ti neighbors are pushed away.  Compared to the pristine material, the nearest Ti--Ti distances increase by 12.3\% across the equatorial vacancy and by 13.4\% across the apical one (9.9\% and 10.1\% for ACE$^+$), in very good agreement to 9.6\% and 8.6\% in DFT~\cite{rusevich_ab_2020}. It should be underlined that the MLIPS thus implicitly cover the repulsion even without the explicit treatment of charges or charge transfer. Both vacancies and the short-range relaxation patterns also remain stable for the tested time interval of 15~ps during MD simulations at 250~K.

The training dataset contained a limited number of planar stacking faults on the $\{110\}$ plane of the cubic phase.  The existence of metastable stacking faults on this plane, corresponding to local minima of the generalized stacking fault surface ($\gamma$-surface), is important for the dislocation behavior in the material. The most relevant planes for oxidic perovskites  are energy profiles along the $\langle 110 \rangle$ and $\langle 100 \rangle$ directions of the $\{110\}$ $\gamma$-surface, which coincide with the directions of the shortest Burgers vectors~\cite{hirel_theoretical_2010}. The profiles calculated by ACE are shown in Fig.~\ref{fig:gs}. The ACE predictions capture correctly the existence of a local minimum along the $[110]$ direction which corresponds to a metastable $1/2[110]$ anti-phase boundary (APB) with an energy of 844~mJ/m$^{2}$ (compared to 929 mJ/m$^{2}$ obtained using DFT with PBE functional~\cite{hirel_theoretical_2010}). Importantly, ACE correctly predicts a shallow local minimum while 
  an  APB energy of only 21~mJ/m$^{2}$ has been reported for core-shell potentials.~\cite{hirel_theoretical_2010} This may open up the possibility for atomistic modelling of $\langle 110\rangle$ dislocations, which was not possible with the core-shell potentials.
  However, it should be noted that despite the good agreement between ACE and DFT, the relaxed stacking fault configurations exhibit rather large values of the extrapolation grade (up to 80). This is a signature that the current ACE parametrization is already in the extrapolative regime for these defect configurations and further validation and upfitting via active learning is required.

\begin{figure}[thb]
\centering
    \begin{center}
 \begin{overpic}[width=0.35\textwidth,clip,trim=0cm .5cm 0cm .5cm]{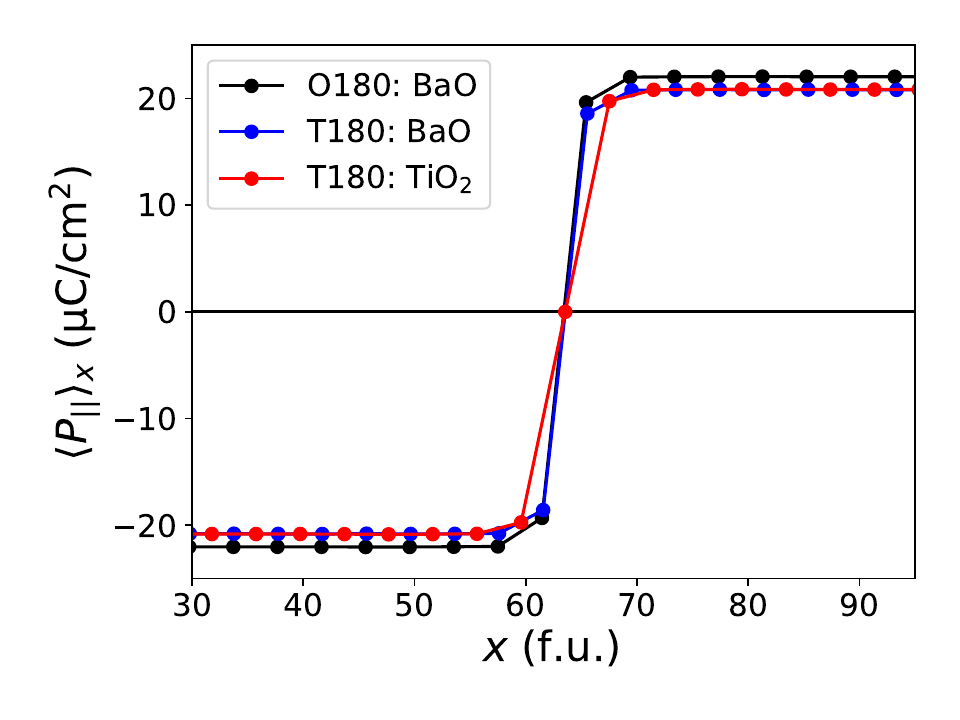}
 \put(3,63){\large{(a)}}
 \end{overpic}
        \begin{overpic}[width=0.35\textwidth,clip,trim=0cm .5cm 0cm .5cm]{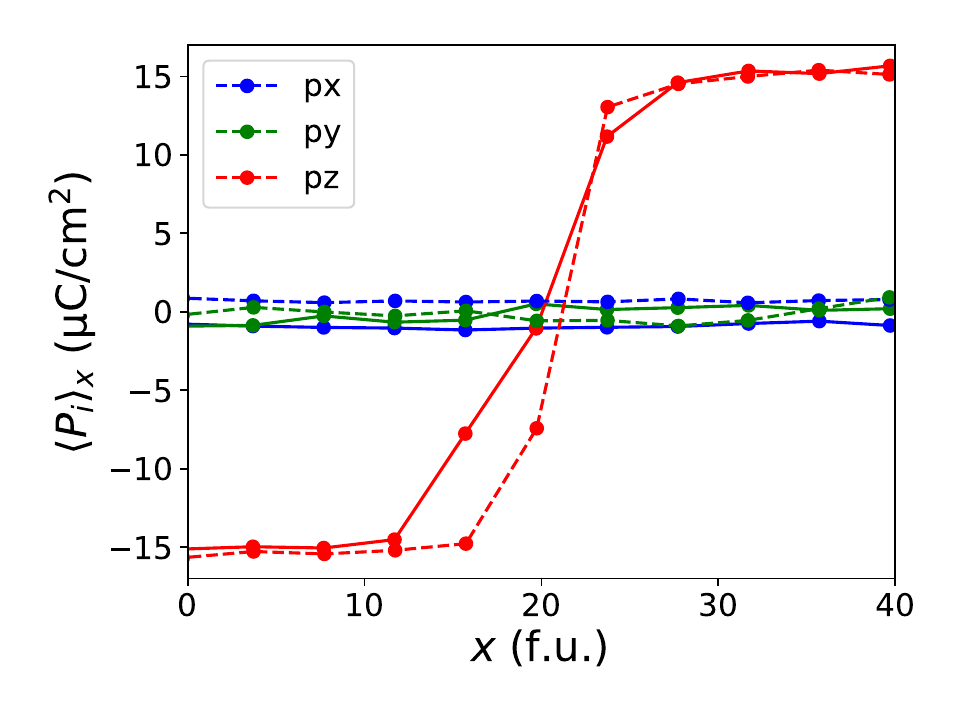} 
\put(3,63){\large{(b)}}
\end{overpic}
 \begin{overpic}[width=0.4\textwidth, clip, trim=-1cm 4.5cm 5cm 0cm ]{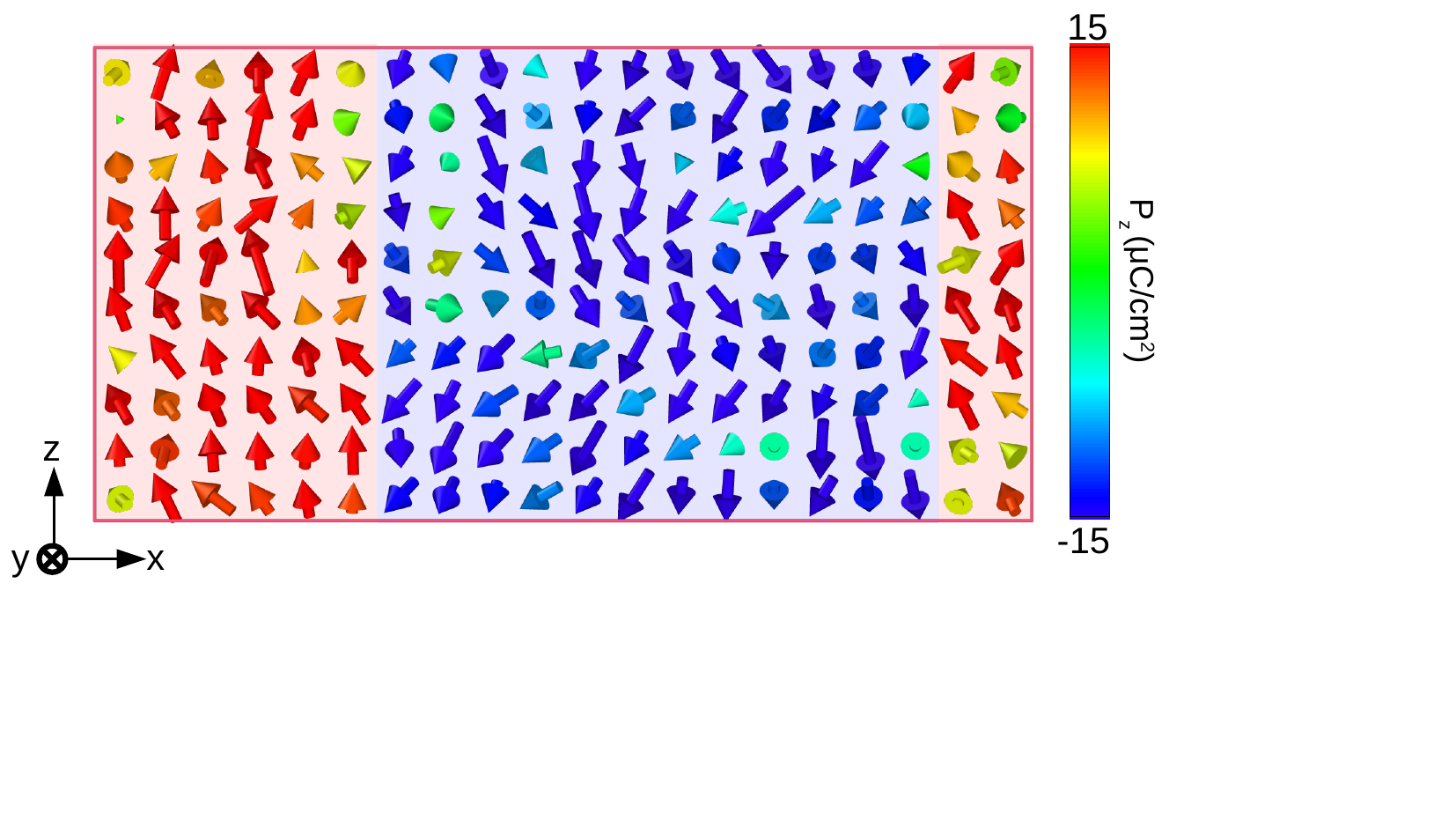}
\put(0,40){\large{(c)}}
\end{overpic}
\end{center}
    \caption{Polarization profiles for (a) T 180$^{\circ}$ and O 180$^{\circ}$ domain walls predicted by ACE at 0~K  and (b) T 180$^{\circ}$ domain wall at 250~K averaged for 5~ps after 45~ps thermalization. The solid and dashed curves refer to the left and right domain walls shown in (c), respectively. In panel (c), the arrows show the local polarization of each unit cell and they are color coded by their P$_z$ component. The red and blue background mark the regions of up and down domain region in the perfect T 180$^{\circ}$ domain wall.}
    \label{fig:T180}
\end{figure}

Figure~\ref{fig:T180}~(a) summarizes the success of the ACE potential to predict properties of domain walls of BTO at 0~K. 
As examples, we have chosen TiO$_2$ and BaO centered walls in the tetragonal phase and the BaO centered walls in the orthorhombic phase. Note that the latter has not been part of the training data. 
In all three cases,  the expected Ising-like switching of the polarization is reproduced (only  $P_{||}$, i.e.\ $\pm P_z$ in T, and $\pm P_yP_z$ in O phase are shown). Following Marton et al.~\cite{marton_domain_2010}, we estimate DW widths of 2.76 and 4.40~\AA~for the BaO and TiO$_2$ centered T 180$^{\circ}$ walls in good agreement to  2.8 and 4.4~\AA~reported in literature \cite{grunebohm_domain_2012}.  Furthermore, we find that the BaO-centered O 180$^{\circ}$ wall is slightly thicker (2.84~\AA), which is again consistent with other studies~\cite{marton_domain_2010,grunebohm_impact_2020}.
In addition, the predicted DW energies of 11~mJ/m$^2$ and 46~mJ/m$^2$ for BaO- and TiO$_2$-centered T 180$^{\circ}$ walls, respectively, are of the same order of magnitude as earlier DFT (LDA) based values of 6 and 14~mJ/m$^2$ \cite{grunebohm_domain_2012}. Note that PBEsol predicts larger energy differences between T and C phases and thus also larger DW energies are expected particularly for the TiO$_2$ domain wall compared to LDA. More importantly, the ACE correctly reproduces the higher energy of the TiO$_2$-centered wall, which determines the energy barrier for domain wall motion and thus thermal stability of domains \cite{klomp_thermal_2022} as well as the critical field strengths to activate wall motion.

ACE also succeeds to predict DW properties at finite temperatures. We examined the stability of T 180$^{\circ}$ walls using a $20 \times 10 \times 10$ supercell. As shown in Fig.~\ref{fig:T180}(b)--(c), the periodic arrangement of parallel domain walls separated by only 10~f.u.\ of bulk crystal remain stable at 250~K. The walls keep the 
expected Ising-like character, while thermal fluctuations induce the expected broadening of the wall. This broadening is related to clusters of reversed polarization on the domain wall, which are critical for DW stability and field-induced motion \cite{klomp_thermal_2022,shin_nucleation_2007}.  
Interestingly, such thin domains were not stable at 250~K, when simulated with the ACE$^+$ potential. Instead, the system switched to the SD tetragonal phase polarized along $x$ (i.e., normal to the initial wall position) during 40~ps.  This result is consistent with the smaller energy barrier for homogeneous polarization switching in ACE$^+$  discussed above. Note that it has been shown that the thermal stability of thin domains depends crucially on the temperature difference to $T_C$ \cite{klomp_thermal_2022} and the ACE$^+$ potential may be applied to study DWs at lower temperatures or for lower DW densities (larger supercells).  Similarly as for the polarization switching, the explicit treatment of charges in the ACE$^+$ potential does not  improve the description of DWs at finite temperatures.


\section{Conclusions}

In summary, we generated two ACE parametrizations for BaTiO$_3$ (BTO) based on a limited dataset of DFT calculations to examine the influence of implicit and explicit treatment of long-range Coulomb interactions. Both models are able to reproduce diverse static properties of BTO, including subtle energy differences between the cubic, tetragonal, orthorhombic and rhombohedral phases as well as their phonon spectra and elastic moduli. Apart from these static properties, the models correctly capture qualitative changes of ferroelectric polarization with temperature including the character of all temperature driven phase transitions, field-induced switching and properties of selected ferroelectric domain walls.

Even though point defects have not been part of the training data, atomic relaxations around oxygen vacancies are reproduced in close agreement with DFT results. The correct description of selected stacking faults and finite temperature domain wall properties makes these potentials promising for studies of dislocations. However, relatively high values of the extrapolation grade show that additional active learning iterations and upfitting are still necessary to improve the transferability of the ACE parametrizations to extended defects. In the same spirit, the correct description of oxygen vacancies is an ideal starting point to explore the validity of different types of point defects. 

 We observed that the stability of the ferroelectric tetragonal phase is reduced when including charges explicitly, and larger thermal fluctuations challenge the stability of the MD simulations. These  findings are in agreement to earlier reports on SiO$_2$, stating that even charge-equilibration models do not improve training errors or phonon spectra, but may lead to worse predictions \cite{Novikov_improving_2019}.  \\
Adding charges explicitly does not seem to improve the description of properties investigated here, while the explicit treatment of Coulomb interactions increases the simulation times by more than a factor of two. The advantage of direct application of electrical field to atoms with explicit charges is not crucial, since the coupling can be modelled reliably also by effective forces.

\section*{Acknowledgements}
AG and LT would like to acknowledge financial support by the German research 
foundation (GR 4792/3, 412303109). We thank Aris Dimou who contributed to the testing of an earlier generation of potentials and Karsten Albe for fruitful discussions.
M.N.P.\ gratefully acknowledges the financial support under the scope of the COMET program within the K2 Center ?Integrated Computational Material, Process and Product Engineering (IC-MPPE)? (Project No 886385).

DFT training data and ACE potential parameters can be obtained from the authors upon reasonable request.

%

\end{document}


\preprint{BTO Supplementary Material}

\title{Efficient local atomic cluster expansion for BaTiO$_3$ close to equilibrium \\ Supplementary Material}
\author{Anna Gr\"unebohm}
\affiliation{Interdisciplinary Centre for Advanced Materials Simulation (ICAMS),  Ruhr-University Bochum, 44780  Bochum, Germany}
\affiliation{Center for Interface-Dominated High Performance Materials (ZGH), Ruhr-University Bochum, 44780  Bochum, Germany}
\author{Matous Mrovec}
\affiliation{Interdisciplinary Centre for Advanced Materials Simulation (ICAMS),  Ruhr-University Bochum, 44780  Bochum, Germany}
\author{Maxim N.\ Popov}
\affiliation{Materials Center Leoben (MCL) Forschung GmbH, A-8700 Leoben, Austria}
\author{Lan-Tien Hsu}
\affiliation{Interdisciplinary Centre for Advanced Materials Simulation (ICAMS),  Ruhr-University Bochum, 44780  Bochum, Germany}
\author{Yury Lysogorskiy} 
\affiliation{Interdisciplinary Centre for Advanced Materials Simulation (ICAMS),  Ruhr-University Bochum, 44780  Bochum, Germany}
\affiliation{Center for Interface-Dominated High Performance Materials (ZGH), Ruhr-University Bochum, 44780  Bochum, Germany}
\author{Anton Bochkarev} 
\affiliation{Interdisciplinary Centre for Advanced Materials Simulation (ICAMS),  Ruhr-University Bochum, 44780  Bochum, Germany}
\author{Ralf Drautz}\affiliation{Interdisciplinary Centre for Advanced Materials Simulation (ICAMS), Ruhr- Ruhr-University Bochum, 44780  Bochum, Germany}
\affiliation{Center for Interface-Dominated High Performance Materials (ZGH), Ruhr-University Bochum, 44780  Bochum, Germany}

\maketitle

\section{Additional validation tests of ACE and ACE$^+$ parametrizations}
\begin{figure}[h]
\centering
\includegraphics[width=0.5\textwidth]{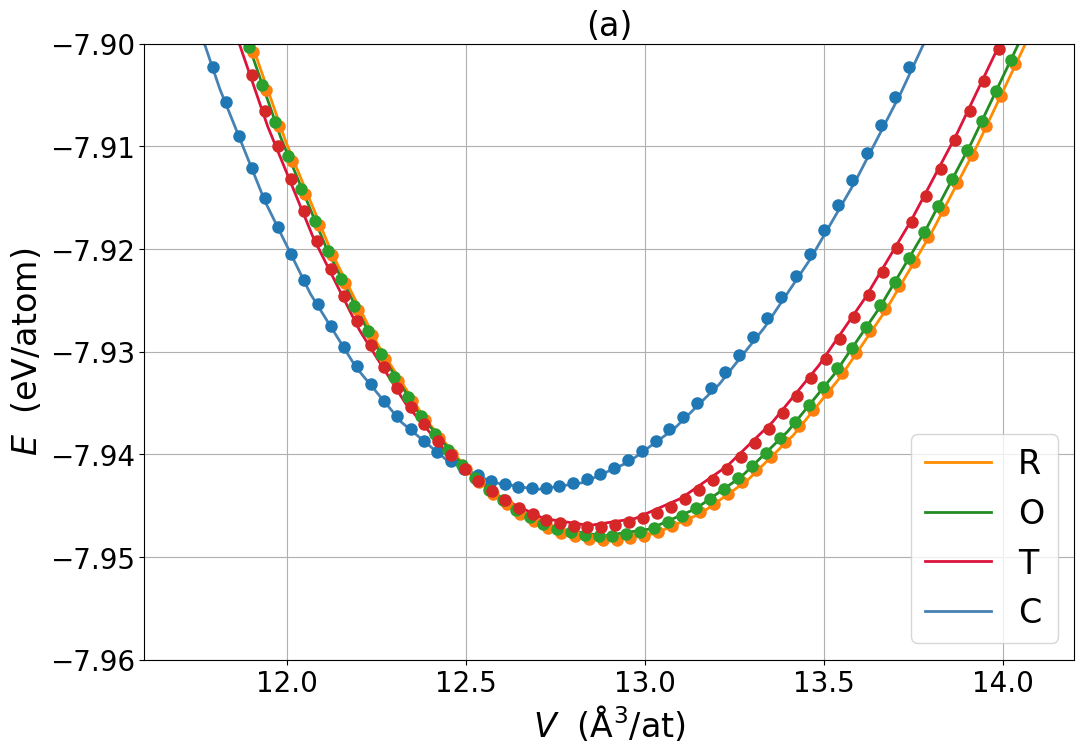}
\vspace{1.0cm}
\centering
\includegraphics[width=0.5\textwidth]{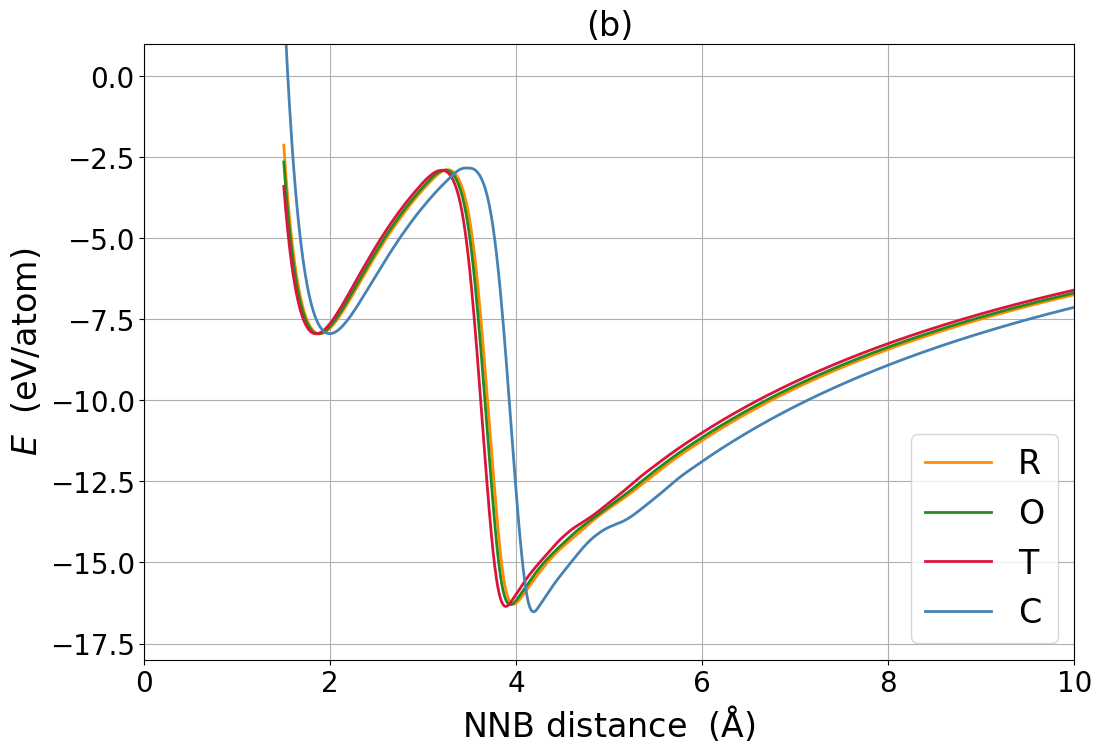}
    \caption{(a) Energy-volume curves for ideal BTO phases predicted by ACE$^{+}$ (lines) together with the reference DFT values (symbols) and (b) Energy vs NNB for ideal BTO phases over broad range of densities predicted by ACE$^{+}$. 
    \label{fig:EVplus}  Panel (b) illustrates the local nature of the ACE$^{+}$ paramerization which can only be used in the vicinity of equilibrium atomic configurations and has no transferability to largely distorted atomic configurations. }
\end{figure}

In this supplementary material we collect additional tests of the ACE$^+$ parameterization.
Figure~\ref{fig:EVplus}~(a) shows the very good agreement of the energy-volume curves of all four phases with DFT data. If superimposed with preditions by ACE, no differences can be seen by the eye.

Figure~\ref{fig:EVplus}~(b) shows that the ACE$^+$ model  cannot predict the limiting case of isolated atoms (NNB distances going to infinity) as expected for a model with fixed ionic charges. Furthermore, while the energy landscape around the equilibrium volume is correctly reproduced, the increase of atomic distances by more than 50\% results in an incorrect minimum of the energy. This shows that large atomic distortions cannot be addressed with ACE$^+$ without further upfitting of the model.

Table~\ref{tab:elast} compares the elastic constants predicted by ACE and ACE$^+$ to our own DFT data. All elastic constants are well reproduced by both types of potentials with a slightly smaller error bar for ACE potentials.
\begin{table}[h]
\centering   
\caption{\label{tab:elast} Comparison of the elastic constants (in GPa) for all four BTO phases obtained by DFT and the ACE potentials. }
\begin{ruledtabular}
\begin{tabular}{lccc}
~  & DFT & ACE & ACE$^{+}$ \\
\hline
\noalign{\vskip 2mm}
Rhombohedral  & & & \\
$C_{11}$  &  145  &  147  & 153 \\ 
$C_{12}$  &  107  &  115  & 101 \\
$C_{44}$  &  116  &  116  & 115 \\
$C_{66}$  &   19  &   16  &  26 \\
\noalign{\vskip 2mm}
\hline
\noalign{\vskip 2mm}
Orthorhombic  & & & \\
$C_{11}$    &  144  &  146  &  149 \\ 
$C_{12}$    &  103  &  110  &   97 \\ 
$C_{13}$    &   98  &  105  &   90 \\ 
$C_{33}$    &  289  &  300  &  300 \\ 
$C_{44}$    &   76  &   78  &  64  \\
$C_{66}$    &  113  &  115  &  110 \\
\noalign{\vskip 2mm}
\hline
\noalign{\vskip 2mm}
Tetragonal & & & \\
$C_{11}$    &  299  &  305  & 299 \\ 
$C_{12}$    &  112  &  122  &  97 \\ 
$C_{13}$    &   90  &   97  &  88 \\ 
$C_{33}$    &  141  &  142  & 150 \\ 
$C_{44}$    &   11  &   19  &   4 \\
$C_{66}$    &  126  &  125  & 124 \\
\hline
\noalign{\vskip 2mm}
Cubic  & & & \\
$C_{11}$    &  316  &  316  & 305 \\ 
$C_{12}$    &  109  &  118  & 107 \\ 
$C_{44}$    &  128  &  125  & 126 \\
\noalign{\vskip 2mm}
\end{tabular}
\end{ruledtabular}
\end{table}

\begin{figure}
    \centering
    \includegraphics[width=0.45\textwidth]{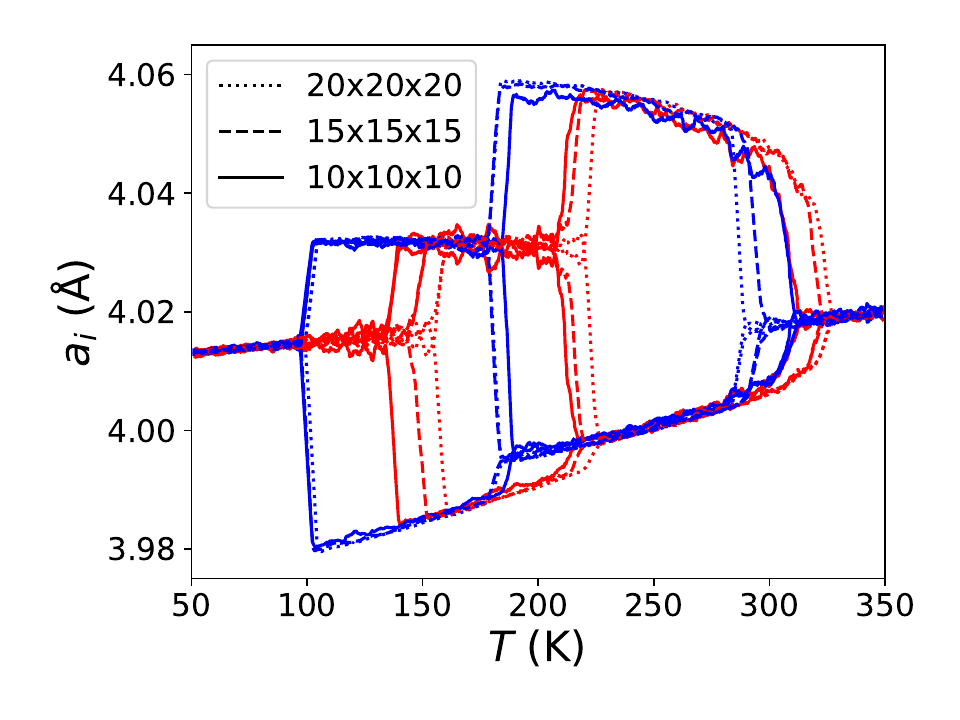}
    \caption{Changes of phase diagram in cooling (blue) and heating (red) simulations with system size (linestyle) as predicted by ACE. 
    \label{fig:cutoff}}
\end{figure}

Figure~\ref{fig:cutoff} shows how the system size in case of ACE potentials influence the phase diagram.  
All phases,  their lattice parameters and polarization (not shown) as well as the intermediate transitions are well reproduced by a minimal setup.
 With system size, the energy barrier for the phase transition of the  material increases. Therefore transition temperatures increase and decrease in heating and cooling, respectively, and the thermal hysteresis increases from 7~K, 18~K, 31~K for 10x10x10 to 31~K, 36~K, 50~K for a system size of $20x20x20$. Despite these quantitative differences, the character of the transition, lattice parameters and polarization are well reproduced for the $10x10x10$ system.

\begin{figure}[htb!]
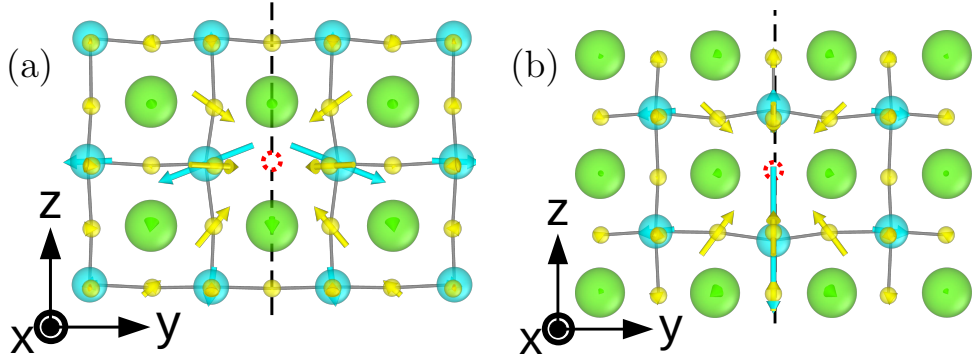

\centering
\begin{overpic}[trim= 6cm 1cm 13.2cm 8cm,clip, angle=180, width=0.4\textwidth]{oxygenvacancy.pdf}
\put(5,60){\large\text{(a)}}
\end{overpic}
\begin{overpic}[trim= 14.8cm 1cm 4cm 8cm,clip, angle=180, width=0.4\textwidth]{oxygenvacancy.pdf}
\put(5,60){\large\text{(b)}}
\end{overpic}
    \caption{Distortions of the atomic structure induced by oxygen vacancies in the tetragonal phase (the tetragonal axis lies along $z$) of BTO as predicted by ACE$^{+}$ at 0~K; (a) the equatorial vacancy connecting Ti atoms perpendicular to the polarization direction and (b) the apical (or axial) vacancy connecting Ti atoms along the polarization direction.
The color coding for atoms and the arrows are the same as Fig. 8.}
    \label{fig:point2}
\end{figure}
Figure~\ref{fig:point2} shows the atomic relaxations around (a) equatorial and (b) apical oxygen vacancies predicted by ACE$^+$. The qualitative differences of atomic relaxation compared to DFT and ACE are discussed in the main paper.

\begin{figure}[ptbhpb]
    \centering
\includegraphics[width=0.5\textwidth]{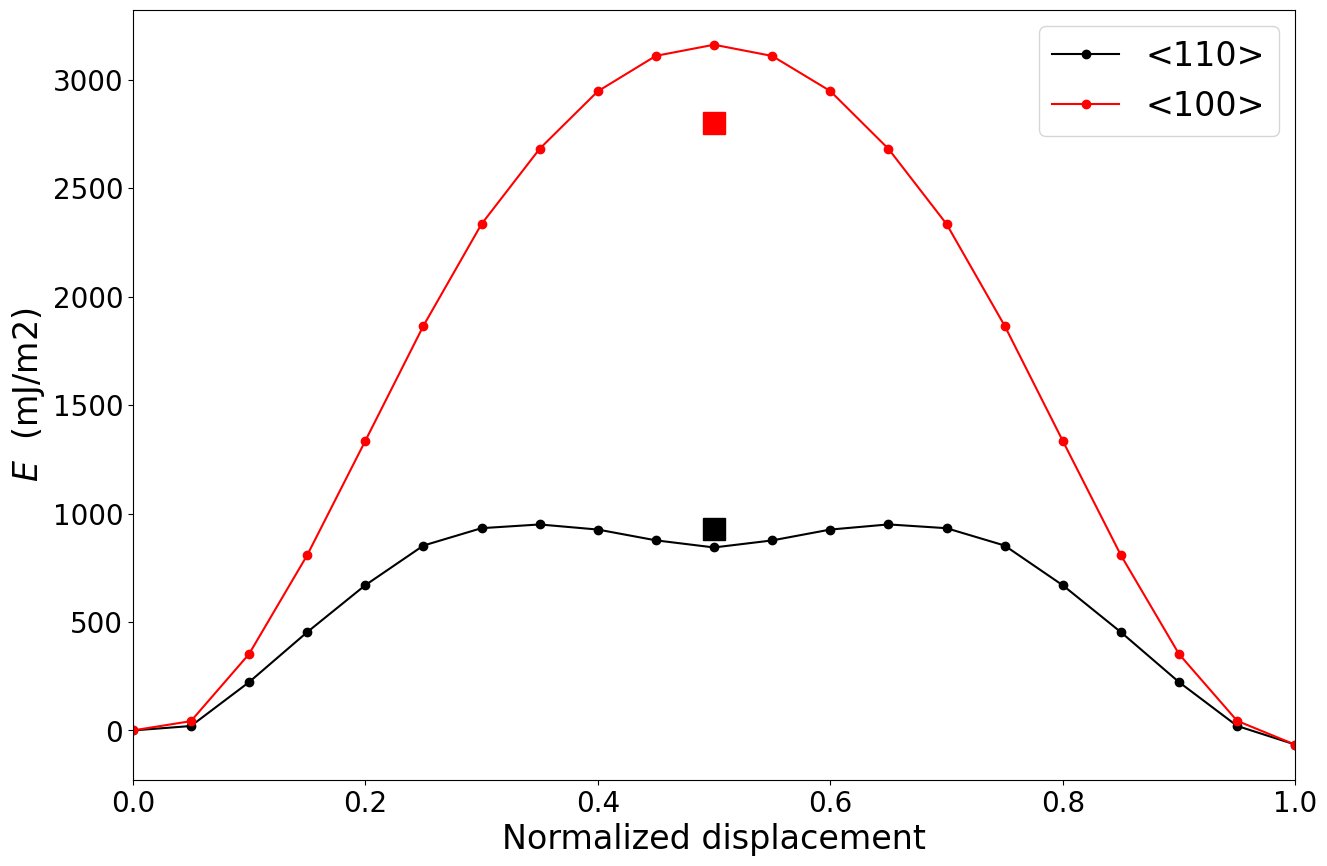}
    \caption{Energy profiles along the $\langle 110 \rangle$ and $\langle 100 \rangle$ directions of the $\{110\}$ $\gamma$-surface computed by ACE$^{+}$. The square symbols are APB energies obtained by PBE DFT~\cite{hirel2010}.}
    \label{fig:gs}
\end{figure}
Figure~\ref{fig:gs} shows the  energy profiles along the $\langle 110 \rangle$ and $\langle 100 \rangle$ directions of the $\{110\}$ $\gamma$-surface computed by ACE$^{+}$, analogous to the results in the main paper based on ACE potentials, the DFT results are well reproduced.

\begin{figure}
    \centering
    \includegraphics[trim=4cm 1cm 6cm 1.3cm,clip,width=0.5\textwidth]{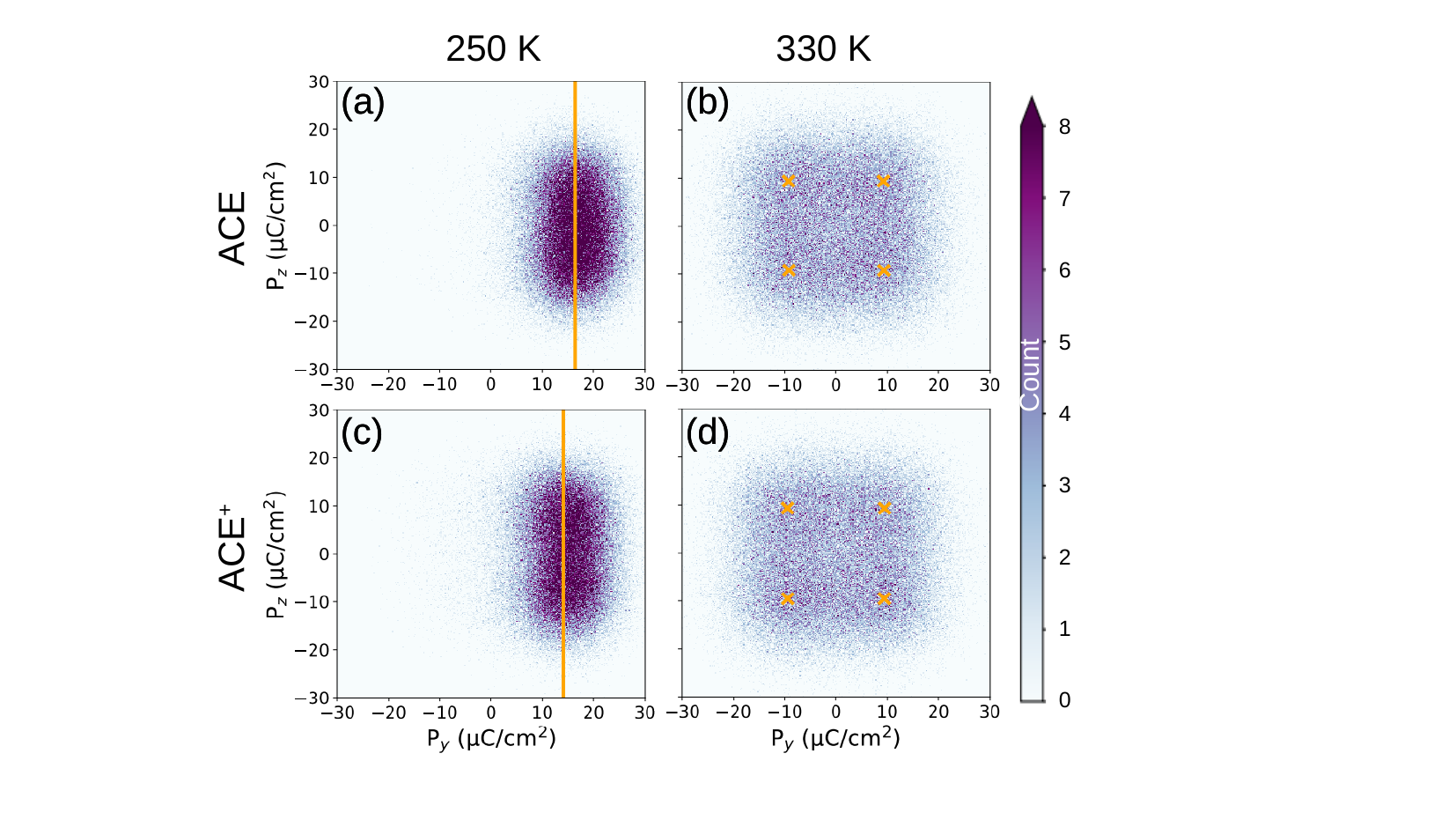}
    \caption{ Cumulative polarization distribution in $P_y-P_z$ space (a, c) in the tetragonal phase at 250~K  and (b, d) in the cubic phase at 330~K during 25~ps (100 snapshots) following the cooling configurations at the respective temperatures for both the ACE and ACE$^+$ potentials. The orange lines and crosses, number of snapshot, bins size, and color bar are the same as Fig. 6. }
    \label{fig:pdist_yz}
\end{figure}
Figure~\ref{fig:pdist_yz} shows the polarization distribution in $P_y$--$P_z$ plane. In accordance to the eight side model, the same distribution as in $P_x$-$P_y$ plane, cf. Fig. 6, is found, both in  C and T phase and the main difference between ACE and ACE$^{+}$ are slightly smaller mean polarization due to larger fluctuations and a shift of the weight of the peak to smaller $P_y$ for the latter.
